\documentclass[12pt,runningheads]{llncs}
\usepackage[margin=1in]{geometry}

\usepackage{booktabs}
\usepackage[linesnumbered,ruled,vlined]{algorithm2e}

\SetAlFnt{\small}
\SetAlCapFnt{\small}
\SetAlCapNameFnt{\small}
\SetAlCapHSkip{0pt}
\IncMargin{-\parindent}

\usepackage[T1]{fontenc}
\usepackage{csquotes}
\usepackage{mathtools}
\usepackage{amsfonts}
\usepackage{upgreek}
\usepackage{amssymb}
\usepackage{booktabs}
\usepackage{csvsimple}

\usepackage{amsmath}
\usepackage{thmtools, thm-restate}
\usepackage{hyperref}
\usepackage{xcolor}
\hypersetup{colorlinks, linktocpage, linkcolor={red!50!black}, citecolor={blue!50!black}, urlcolor={blue!80!black}}
\usepackage[capitalise]{cleveref}
\usepackage{xparse}
\usepackage{bbm}
\usepackage{dsfont}

\usepackage{colorblind}
\usepackage{pgfplots}
\pgfplotsset{compat=newest}
\usetikzlibrary{dateplot}
\newcommand*\shortyear[1]{\expandafter\@gobbletwo\number\numexpr#1\relax}
\usetikzlibrary{patterns}
\usetikzlibrary{arrows.meta, fit, decorations.pathreplacing, calligraphy, math, arrows, shapes.arrows, shapes.geometric}
\definecolor{darkkhaki204185116}{RGB}{204,185,116}
\definecolor{darkslategray38}{RGB}{38,38,38}
\definecolor{indianred1967882}{RGB}{196,78,82}
\definecolor{lightgray204}{RGB}{204,204,204}
\definecolor{lightslategray129114178}{RGB}{129,114,178}
\definecolor{mediumseagreen85168104}{RGB}{85,168,104}
\definecolor{steelblue76114176}{RGB}{76,114,176}

\crefname{definition}{Definition}{Definitions}
\crefname{theorem}{Theorem}{Theorems}
\crefname{corollary}{Corollary}{Corollaries}
\crefname{example}{Example}{Examples}
\crefname{remark}{Remark}{Remarks}

\DeclareMathOperator*{\argmax}{arg\,\max}

\newcommand{\define}{\stackrel{\mathclap{\mbox{\text{\tiny def}}}}{=}}

\usepackage[symbols,acronym,nonumberlist,nogroupskip,numberedsection,stylemods={mcols,longbooktabs}]{glossaries-extra}
\glssetcategoryattribute{acronym}{nohyper}{true}
\setabbreviationstyle[acronym]{long-short}
\newacronym{PDF}{PDF}{probability density function}
\newacronym{CDF}{CDF}{cumulative density function}
\newacronym{iid}{i.i.d.}{independent and identically distributed}
\newacronym{wrt}{w.r.t.}{with regard to}
\newacronym{wlog}{w.l.o.g.}{without loss of generality}
\newacronym{TTL}{TTL}{time to live}
\newacronym{QoS}{QoS}{quality of service}
\newacronym{FIFO}{FIFO}{first-in-first-out}

\glsxtrnewsymbol[description={A transaction.}]{tx}{\ensuremath{\tau}}
\newcommand{\tx}{{\gls[hyper=false]{tx}}}

\glsxtrnewsymbol[description={\gls[hyper=false]{TTL} of a transaction.}]{ttl}{\ensuremath{t}}
\newcommand{\ttl}{{\gls[hyper=false]{ttl}}}

\glsxtrnewsymbol[description={\gls[hyper=false]{TTL} of past transactions.}]{pending}{\ensuremath{\mu}}
\newcommand{\pending}{{\gls[hyper=false]{pending}}}

\glsxtrnewsymbol[description={Immediacy ratio for our non-myopic allocation rule.}]{imratio}{\ensuremath{\ell}}
\newcommand{\imratio}{{\gls[hyper=false]{imratio}}}

\glsxtrnewsymbol[description={Transaction schedule function.}]{adversary}{\ensuremath{\psi}}
\newcommand{\adversary}{{\gls[hyper=false]{adversary}}}

\glsxtrnewsymbol[description={Allocation function.}]{allocation}{\ensuremath{x}}
\newcommand{\alloc}{{\gls[hyper=false]{allocation}}}

\glsxtrnewsymbol[description={Transaction fee.}]{fee}{\ensuremath{\phi}}
\newcommand{\fee}{{\gls[hyper=false]{fee}}}

\glsxtrnewsymbol[description={Maximal block-size.}]{blocksize}{\ensuremath{\mathcal{B}}}
\newcommand{\blocksize}{{\gls[hyper=false]{blocksize}}}

\glsxtrnewsymbol[description={The discount factor.}]{discount}{\ensuremath{\lambda}}
\newcommand{\discount}{{\gls[hyper=false]{discount}}}

\glsxtrnewsymbol[description={Planning horizon.}]{horizon}{\ensuremath{T}}
\newcommand{\horizon}{{\gls[hyper=false]{horizon}}}

\glsxtrnewsymbol[description={Revenue.}]{revenue}{\ensuremath{u}}
\newcommand{\revenue}{{\gls[hyper=false]{revenue}}}

\newcommand{\ttlDomain}{\ensuremath{\mathcal{T}}}
\newcommand{\feeDomain}{\ensuremath{\Phi}}
\newcommand{\txDomain}{\ensuremath{\ttlDomain \times \feeDomain}}

\NewDocumentCommand{\expect}{ e{_} s o >{\SplitArgument{1}{|}}m }{%
  \operatorname{E}%     the expectation operator
  \IfValueT{#1}{{\!}_{#1}}% the measure of the expectation
  \IfBooleanTF{#2}{% *-variant
    \expectarg*{\expectvar#4}%
  }{% no *-variant
    \IfNoValueTF{#3}{% no optional argument
      \expectarg{\expectvar#4}%
    }{% optional argument
      \expectarg[#3]{\expectvar#4}%
    }%
  }%
}
\NewDocumentCommand{\expectvar}{mm}{%
  #1\IfValueT{#2}{\nonscript\;\delimsize\vert\nonscript\;#2}%
}
\DeclarePairedDelimiterX{\expectarg}[1]{[}{]}{#1}

\allowdisplaybreaks

\begin{document}
\title{Online Packet Scheduling With Time Discounts}
\author{Yotam Gafni\inst{1}\orcidID{0000-0002-2144-655X} \and
Aviv Yaish\inst{2}\orcidID{0000-0002-7971-2494}}
\authorrunning{Y. Gafni and A. Yaish} 

\institute{Weizmann Institute of Science \\
\email{yotam.gafni@gmail.com} \and
Yale University, IC3, Complexity Science Hub Vienna \\
\email{a@yai.sh}}

\maketitle

\begin{abstract}
We study a \emph{financial} version of the online problem of scheduling weighted packets with deadlines.
The main novelty is that, while prior works assume packets have \emph{fixed} weights, we consider packets with \emph{time-decaying} values.
Such considerations are natural in financial environments, where the present value of future actions may be discounted.

We analyze the competitive ratios of scheduling algorithms under a range of discount rates encompassing the traditional undiscounted case where weights are fixed (i.e., a discount rate of 1), the fully discounted myopic case (i.e., a rate of 0), and those in between.
We show how existing methods from the literature perform suboptimally in the more general discounted setting.
Notably, we devise a novel memoryless deterministic algorithm, and prove that for discount factors up to $\approx 0.77$, it guarantees the best competitive ratio attainable by deterministic algorithms.
Moreover, we develop a randomized algorithm and prove that it outperforms the best possible deterministic algorithm for any discount rate.

\end{abstract}

\section{Introduction}
We study an online problem where an algorithm is tasked with allocating incoming weighted packets with deadlines, with the goal of maximizing the total weight of allocated packets, with the double admonition that each packet expires after a given deadline, and even before expiry, its value decays over time.
The difficulty of the problem arises when the number of packets that can be concurrently allocated is limited, and no information about future incoming packets is given in advance.
The undiscounted problem where values do not decay is considered fundamental \cite{vesely2021packet,tsanikidis2023optimal,liang2024learning}, and is widely applied to communication networks \cite{hajek2001competitiveness,tsanikidis2023optimal}, ride sharing \cite{ashlagi2019edge,dickerson2021allocation}, and time-sensitive decisions \cite{vesely2018online}, among other settings.

Despite some of the problem's applications being rooted in financial motivations \cite{vesely2018online,ashlagi2019edge,dickerson2021allocation}, the problem's classic model considers packets with \emph{static} weights, and thus does not capture the \emph{time value of money} \cite{fisher1930theory}.
This common financial consideration emphasizes that one may prefer to receive a fixed amount of funds sooner rather than later, as it can be invested and thus increase in value \cite{fisher1911purchasing}.
This insight extends to other settings, e.g., consider perishable commodities such as food and pharmaceuticals which have hard expiration dates and also naturally decay over time, whether in terms of quality or quantity \cite{van1963inventory,nahmias1982}.

These considerations, which can be formalized as the \emph{discounting} of future utility by a factor, add a novel dimension of complexity that a scheduling algorithm has to face:
respecting incoming packets' ``hard'' deadline constraints may not suffice to achieve optimality.
This is because because packets can change in weight even beforehand, as dependent on the discount rate.
Given the increased difficulty, a natural question arises:
\begin{quote}
    \emph{Can we devise scheduling algorithms with good performance guarantees that account for the time value of money, for a variety of discount rates?}
\end{quote}

\subsection{Our Contributions}
In our work, we provide an affirmative answer to this question by presenting two scheduling algorithms which discount the value of future actions according to a given factor $\discount$:
the $\imratio$-immediacy-biased ($\imratio IB$) algorithm which attains the optimal deterministic performance for any discount rate $\discount \in \left[0, \approx 0.77\right]$, and the randomized $RDISC$ algorithm which provides guarantees beyond those given by the best possible deterministic algorithm for any $\discount \in \left[0,1\right]$.

\begin{figure}
    \centering
    \scalebox{1.0}{\begin{tikzpicture}[
declare function={
upper(\x) =
(\x < 0.99) * min(
    (2 + x * (x + sqrt(4 + x*x))) /
    (2 + x * (2 + x * (3*x + sqrt(4 + x*x))))
    ,
    (2/(\x + sqrt(\x^2 + 4)))
) + and(\x==0.99, \x < 1) * (0.570537) + (\x==1) * (0.5)
;}]
    \begin{axis}[
        cycle list={{mediumseagreen85168104}, {steelblue76114176}, {indianred1967882}, {lightslategray129114178}, {darkkhaki204185116}, {darkslategray38},},
        domain=0:1,
        xmin=0, xmax=1, ymin=0, ymax=1,
        xmajorgrids, ymajorgrids, xmajorticks, ymajorticks,
        samples=100,
        xlabel = Discount Rate,
        ylabel = Competitive Ratio,
        axis line style={lightgray204},
        legend style={
          font=\tiny,
          anchor=west,
          at={(1.02,0.5)},
          fill opacity=0.8,
          draw=lightgray204,
        },
        tick align=outside,
        tick style={color=white},
        every axis plot/.append style={thick},
    ]
        \addplot+[dashed,color=red] {1 - x/4};
        \addplot+[dashed,color=blue] coordinates {
            (0.001, 0.9995001666249781) (0.01, 0.9950166250831893) (0.02, 0.9900663346622374) (0.03, 0.9851488817163949) (0.04, 0.9802640211919206) (0.05, 0.9754115099857197) (0.060000000000000005, 0.9705911069291879) (0.06999999999999999, 0.9658025727721676) (0.08, 0.961045670167053) (0.09, 0.9563201636530202) (0.09999999999999999, 0.9516258196404038) (0.11, 0.946962406395198) (0.12, 0.9423296940236877) (0.13, 0.9377274544572207) (0.14, 0.9331554614371009) (0.15000000000000002, 0.9286134904996145) (0.16, 0.9241013189611791) (0.17, 0.9196187259036253) (0.18000000000000002, 0.9151654921595999) (0.19, 0.9107414002980933) (0.2, 0.9063462346100909) (0.21000000000000002, 0.9019797810943472) (0.22, 0.8976418274432796) (0.23, 0.8933321630289823) (0.24000000000000002, 0.8890505788893609) (0.25, 0.8847968677143805) (0.26, 0.8805708238324375) (0.27, 0.8763722431968402) (0.28, 0.872200923372409) (0.29000000000000004, 0.868056663522189) (0.3, 0.8639392643942738) (0.31, 0.8598485283087444) (0.32, 0.8557842591447158) (0.33, 0.8517462623274964) (0.34, 0.8477343448158539) (0.35000000000000003, 0.8437483150893901) (0.36000000000000004, 0.8397879831360249) (0.37, 0.8358531604395819) (0.38, 0.8319436599674846) (0.39, 0.8280592961585521) (0.4, 0.8241998849109017) (0.41000000000000003, 0.8203652435699527) (0.42000000000000004, 0.8165551909165315) (0.43, 0.8127695471550779) (0.44, 0.8090081339019514) (0.45, 0.805270774173837) (0.46, 0.801557292376248) (0.47000000000000003, 0.797867514292126) (0.48000000000000004, 0.7942012670705398) (0.49, 0.7905583792154773) (0.5, 0.7869386805747332) (0.51, 0.7833420023288904) (0.52, 0.7797681769803954) (0.53, 0.776217038342726) (0.54, 0.7726884215296488) (0.55, 0.7691821629445696) (0.56, 0.7656981002699734) (0.5700000000000001, 0.7622360724569524) (0.5800000000000001, 0.758795919714824) (0.59, 0.7553774835008354) (0.6, 0.7519806065099559) (0.61, 0.7486051326647545) (0.62, 0.7452509071053638) (0.63, 0.7419177761795283) (0.64, 0.7386055874327366) (0.65, 0.7353141895984369) (0.66, 0.7320434325883346) (0.67, 0.7287931674827725) (0.68, 0.7255632465211918) (0.6900000000000001, 0.7223535230926732) (0.7000000000000001, 0.719163851726558) (0.7100000000000001, 0.7159940880831477) (0.72, 0.7128440889444837) (0.73, 0.7097137122052021) (0.74, 0.7066028168634669) (0.75, 0.7035112630119804) (0.76, 0.7004389118290668) (0.77, 0.6973856255698336) (0.78, 0.6943512675574057) (0.79, 0.691335702174233) (0.8, 0.688338794853473) (0.81, 0.6853604120704431) (0.8200000000000001, 0.6824004213341472) (0.8300000000000001, 0.6794586911788728) (0.8400000000000001, 0.6765350911558574) (0.85, 0.6736294918250274) (0.86, 0.6707417647468037) (0.87, 0.6678717824739782) (0.88, 0.6650194185436575) (0.89, 0.6621845474692747) (0.9, 0.6593670447326676) (0.91, 0.6565667867762242) (0.92, 0.6537836509950934) (0.93, 0.6510175157294612) (0.9400000000000001, 0.6482682602568924) (0.9500000000000001, 0.6455357647847355) (0.9600000000000001, 0.6428199104425917) (0.97, 0.6401205792748466) (0.98, 0.6374376542332657) (0.99, 0.6347710191696508) (1.0, 0.6321205588285577) };
        \addplot+[color=red] {(x + sqrt(4 +x^2)) / (2 + x^2 + x * sqrt(4 + x^2))};
        \addplot+[color=red, samples at = {0, 0.01, ..., 0.97, 0.98, 0.99, 1}] {upper(x)};
        \addplot+[color=blue] {min(2/(x + sqrt(x*x+4)), 1/(1+pow(x,3)))};
        \addplot+[color=black] {1/(1+x)};
        \addlegendentry{Randomized upper bound (\cref{res:RandomizedUpperBound})}
        \addlegendentry{Randomized lower bound (\cref{res:RDISC})} 
        \addlegendentry{Deterministic upper bound (\cref{res:DeterministicUpperBound})}
        \addlegendentry{$R_{{\imratio}IB}$ upper bound (\cref{res:GoldenUpper})}
        \addlegendentry{$R_{{\imratio}IB}$ lower bound (\cref{res:GoldenGreedy})}
        \addlegendentry{$R_{Greedy}$ tight bound (\cref{res:GreedyCompRatio})}
    \end{axis}
\end{tikzpicture}}
    \caption{Our competitive ratio bounds for the deterministic algorithms we present in \cref{sec:Greedy,sec:Deterministic}, and the randomized ones we give in \cref{sec:Randomized}, for discount rates $\discount \in \left[0,1\right]$.
    In comparison, the work of \cite{vesely2019a} shows that the tight bound for the undiscounted case is $\phi^{-1} \approx 0.618$.
    We use dashed and full lines to respectively represent the randomized and deterministic cases, while upper and lower bounds are respectively colored in red and blue.
    The deterministic lower and upper bound are identical for $\lambda \leq 0.77$.}
    \label{fig:CompetitiveRatios}
\end{figure}

\subsubsection*{Model, Technique, and Results}
We advance a novel packet scheduling setting which captures time preferences by discounting future utility.
With every passing time step, each packet's weight decays by a factor $\discount < 1$, until the packet's deadline passes.
Our key benchmark to evaluate a scheduling algorithm is its \emph{competitive ratio} \cite{borodin2005online}, formalized in \cref{def:CompRatio} as the ratio of the algorithm's performance divided by that of the optimal allocation.\footnote{This implies that $R \in \left[0,1\right]$ and follows the classic work of \cite{hajek2001competitiveness}, while other work may use the inverse ratio, such as \cite{vesely2019a}. 
Since both approaches are used by important preceding work and only differ aesthetically (i.e., the lower bounds of the latter are the upper bounds of the former, and vice versa), we choose the former.}

Due to the discounted setting, we develop a novel method to analyze the performance of scheduling algorithms, which we call the  ``reverse subchain'' technique.
Intuitively, within the set of all decisions made by an algorithm $ALG$, we mark the cases where it was (locally, at a given timestamp) outperformed by the optimal algorithm $OPT$ by over some factor $\imratio$.
In these cases, as long as $ALG$ acts greedily in some sense, it must be that it has already allocated the transaction in a previous step.
This naturally creates ``subchains'' of corresponding transactions.
We isolate the analysis to each such subchain.
Within a subchain, if there is a large number of steps between the first and second transactions, the discount rate guarantees that $ALG$ outperforms OPT over this subchain (as it allocates the large transaction early).
If there is a small number of steps, then we can perform a case analysis.
For example, in our \cref{res:GoldenGreedy}, we can use this method to limit the case analysis to at most a $2$-step gap. 

Our main results, which we also summarized in \cref{fig:CompetitiveRatios}, are as follows:
\begin{itemize}
\item \emph{Bounds for the discounted setting.}
We upper bound the competitive ratios of both deterministic algorithms (in \cref{res:DeterministicUpperBound}) and randomized algorithms (in \cref{res:RandomizedUpperBound}), as dependent on the discount rate.
In particular, these bounds can be seen as parameterized generalizations of \cite{hajek2001competitiveness} and \cite{bienkowski2011randomized} respectively.
However, choosing a meaningful parameterization of the constructions, and extending the limit arguments for a general discount parameter requires careful attention.
For $RDISC_{\discount}$, it requires a restatement of the potential argument condition, and a good choice of sampling distribution (for more details, see \cref{sec:RelatedWork} and the proof overviews of \cref{res:DeterministicUpperBound,res:RandomizedUpperBound}).

\item \emph{A new deterministic algorithm.}
In \cref{def:ImmediacyBiased}, we present our novel $\imratio$-immediacy-biased ($\imratio IB$) algorithm.
When setting $\imratio$ correctly, we prove in \cref{res:GoldenGreedy} that the algorithm matches our deterministic upper-bound: its performance is optimal for $0 \leq \discount \lesssim 0.770018$.
We call this range of discount values the ``semi-myopic'' regime.
\end{itemize}

\subsubsection*{A General Lesson}
We believe that there is a general lesson to be learned from our work which can be applied to other online problems.
An algorithm's competitive ratio is generally driven by its performance given sequences designed to trick it into making the wrong choices. We observe that discounting is not only \textit{financially} natural, but also helps technically by allowing us, in effect, to disregard adversaries that use long and complex bad sequences. 
When the discount is strong enough, simple algorithms may guarantee optimal results. This is clear in the extreme myopic case (where the future is completely discounted, and the greedy algorithm is optimal), but as we show, it also applies for the semi-myopic case, which comprise a substantial portion of the range of discount values. 

\subsection{Related Work}
\label{sec:RelatedWork}

A rich literature studies a closely-related question to the one we formulate, that of packet scheduling.
Moreover, its focus is as ours on using the competitive ratio method to evaluate possible algorithms.
Hajek~\cite{hajek2001competitiveness} shows that greedily choosing the heaviest packet (in our terms: the highest-fee transaction) guarantees a competitive ratio of $\frac{1}{2}$.
We obtain a more general result in \cref{res:GreedyCompRatio}, and show that given discount factor $\discount$, it achieves a competitive ratio of $\frac{1}{1 + \discount}$.
Notice that with $\discount = 0$, i.e., in the myopic case where only the current step matters, choosing greedily becomes optimal, as expected.
Moreover, Hajek~\cite{hajek2001competitiveness} proves an optimal deterministic bound of $\frac{1}{\phi} \approx 0.618$.
We identify that the key property of the golden ratio that makes it ``work'' as the upper bound is that it satisfies: $\frac{\phi}{1+\phi}= \frac{1}{\phi}$, and find that we can develop a version of this equation parameterized by the discount factor $\discount$: $\frac{r}{1 + \discount \cdot r} = \frac{1}{r}$, and solve it as $r = \frac{1}{2}\left(\discount + \sqrt{\discount^2 + 4}\right)$.
We show that this yields an optimal bound in the discounted case for a large regime of parameters.

The packet scheduling literature continued by presenting optimal algorithms for both specific sub-cases (e.g., when deadlines are ``agreeable'') \cite{li2005optimal}, and also for the general undiscounted case \cite{vesely2019a}.
Veselý et al.~\cite{vesely2019a} use a memoryless algorithm as a stepping stone to designing an optimal deterministic algorithm which is not memoryless.
In both cases, the memoryless algorithm does not achieve the optimal deterministic guarantee.
It is thus quite surprising that we are able to show a different result in the discounted case, where the $\imratio IB$ algorithm achieves optimal performance for a large regime of discount factors $\discount$, which we call the semi-myopic regime of $\discount \lesssim 0.770018$.

It is important to note that it may be suboptimal to apply the optimal undiscounted algorithm, PlanM \cite{vesely2019a}, to the discounted case.
To illustrate, our $\imratio IB$ algorithm outperforms PlanM for discount rates $\discount < \phi^{-1} \approx 0.618$.
Let $x\in(\frac{1}{1-\lambda},\phi^2)$ and consider two transactions: one with deadline $1$ and value $1$, and another with deadline $2$ and value $x$.
PlanM first chooses the earlier deadline packet, and then the other, for a total value with discounting of $1 + x \lambda$.
Our $\imratio IB$ (with our optimized choice of $\imratio$) chooses the higher value transaction immediately, for a total discounted value of $x$, which is better.  

The literature also considered randomized algorithms and upper bounds.
Chin et al.~\cite{chin2006online} introduce RMIX, a randomized algorithm that uses a randomized coefficient to determine which packet to allocate.
Bienkowski, Chrobak, and Jeż~\cite{bienkowski2011randomized} extend the analysis of $RMIX$ from the oblivious adversary model to the adaptive adversary one, and also provide an upper bound for any randomized algorithm facing an adaptive adversary.

When moving to our new discounted setting, prior analyses of randomized algorithms require a new: (i) algorithm, (ii) potential function argument for the lower bound, and (iii) adversary and analysis for the upper bound.
First, with respect to the algorithm, RMIX under-prioritizes high-value transactions for $\discount<1$.
Consider transactions $(1,e+1),(2,e^2)$, and $\discount=\frac{1}{e}$.
RMIX has a positive probability of obtaining utility $2e+1=e+1+(e\cdot\frac{1}{e})$ by choosing $(1,e+1)$ first.
RDISC always chooses $(2,e^2)$ first and obtains $e^2>2e+1$.
Second, \cite{chin2006online}'s potential function argument cannot be used as-is because $\discount<1$ affects how the potential changes between steps, and also due to our new sampling distribution.
Third, our adversary is novel.
To make the analysis tractable, we exponentially increase transaction values to allow ignoring leftover transactions.
Intuitively, given transactions $(1,1),(2,b)$, if an algorithm chooses $(1,1)$ the adversary makes this suboptimal by sending two transactions with exponentially larger values; otherwise, the adversary stops.
Deriving the bound requires an intricate analysis: setting $b$ and the next rounds' values is involved and depends on the algorithm's probabilities of choosing each transaction.

Notably, Veselý~\cite{vesely2021packet} highlights that tightening the gap between the upper and lower bounds for randomized algorithms in the undiscounted case as important, and Veselý et al.~\cite{vesely2019a} conclude by saying that establishing tight bounds is the most prominent open problem in online packet scheduling.
We contribute to solving this: we tighten the gap for $\discount\rightarrow$$0$.

\section{Model}
\label{sec:Model}
Our game proceeds in $\horizon \in \mathbb{N}$ rounds.
$T$ is unknown to the allocator. 
At each round, a set of new transactions is broadcast to the network by users.
The allocator has the sole authority to decide which transactions to allocate, among the set of non-expired transactions that were not previously allocated. At each timestamp, at most a single transaction can be allocated. 
A transaction $\tx = \left( \ttl, \fee \right)$ is defined by its \gls{TTL} $\ttl \in \ttlDomain \define \mathbb{N} \cup \left \{ \infty \right \}$ and fee $\fee \in \feeDomain \define \mathbb{R_+}$.
The \gls{TTL} represents a transaction's validity period during which it is eligible for allocation, while its fee is its value for the allocator if allocated immediately upon arrival. 
Otherwise, at any timestamp the fee decays by  a \emph{discount factor} $\discount \le 1$.
We denote $\tx$'s \gls{TTL} and fee by $\ttl(\tx), \fee(\tx)$, respectively.

The transactions broadcast to the network at each turn are given by $\adversary:\{0,\ldots, \horizon\} \rightarrow (\txDomain)^*$, where $(\txDomain)^*$ denotes some amount of transactions in $\txDomain$.
At the $i$-th turn, future arrival for any $j \in \left[i+1,\horizon\right]$ is unknown to the allocator.

The allocation rule $\alloc$ defines the mechanism's allocation of transactions to the upcoming block.
It is possible that $\alloc$ is randomized; in that case, it maps to $\Delta\left(\txDomain\right)$.
We denote the transaction allocated in the $j$-th turn by $\overline{\alloc_j}$.

Valid transactions that were not yet allocated to a block, nor expired, are stored in a data structure called the \emph{pending set}.
In \cref{def:Pending}, we formalize the pending set as a function which receives the current turn as input, and outputs the set of transactions viable for inclusion.
Our definition relies on a useful notion of ``leftover'' transactions which are still unallocated and thus are carried over to the next round, with the pending set function decreasing their \gls{TTL} correspondingly, and then pruning those which expired.
\begin{definition}[Leftover Transactions and the Pending Set]
    \label{def:Pending}
    The pending set function $\pending:\{0, \ldots, T\} \rightarrow (\txDomain)^*$ is defined recursively, starting initially from $\pending(0) \define \emptyset$.
    Given adversary $\adversary$, round $j$ and the corresponding pending set $\pending(j)$, we denote the set $l(j+1)$ of leftover transactions remaining after the allocation of the current round $\overline{\alloc_j}$ as:
    $$
        l(j+1) \define \left( \pending \left(j\right) \cup \adversary\left(j\right) \right) \setminus \overline{\alloc_j}
        .
    $$
    After decreasing the \gls{TTL} of each leftover transaction in $l(j+1)$, the pending set for the next turn $\pending(j+1)$ is defined as the set of transactions which have not expired yet:
    $$
        \pending(j+1) \define \{(\ttl-1, f) | \ttl > 1\}_{(\ttl,f) \in l(j+1)}
    .
    $$
\end{definition}

Given adversarial arrival $\adversary$, the utility function of using an allocation algorithm $\alloc$ is denoted by $\revenue(\alloc | \adversary)$.
We first formalize the utility for a single step in \cref{def:SingleRoundUtility}, and then extend it to all steps in \cref{def:TotalUtility}.
We follow with \cref{eq:detUtility}, which illustrates these definitions by giving a concrete instantiation of the utility function.

\begin{definition}[Single Round Utility]
    \label{def:SingleRoundUtility}
    Let $\pending(j)$ be the set of pending packets at turn $j$, after accounting for the allocations made until that point $\overline{\alloc_1}, \dots, \overline{\alloc_{j-1}}$, and the adversarial arrival thus far.
    Given packet $i$ is allocated in the $j$-th turn, denote its transaction by $\fee_i$, and the fees of all other transactions by $\fee_{-i}$.
    The expected utility of using $\alloc$ at the $j$-th turn is:
    $$
    \revenue_j (\alloc \,\lvert\, \adversary, \overline{\alloc_1}, \dots, \overline{\alloc_{j-1}})
    \define
    \expect*{
    \fee_{\alloc(\adversary(j) \cup \pending(j))}
    \,\lvert\, \adversary, \overline{\alloc_1}, \dots, \overline{\alloc_{j-1}}
    },
    $$
    where the expectation is taken over the randomness used, if the allocation is randomized.
\end{definition}
\begin{definition}[Total Utility]
    \label{def:TotalUtility}
    For all $j$, denote the allocation made at round $j$ by $\overline{\alloc_j}$.
    Given discount factor $\discount$, the expected utility until round $\horizon$ is:
    $$\revenue (\alloc \,\lvert\, \adversary) \define \expect*{\sum_{j=0}^\horizon  \discount^j \revenue_j (\alloc \,\lvert\, \adversary, \overline{\alloc_1}, \dots, \overline{\alloc_{j-1}})},$$
    where the expectation is taken over the randomness of $\overline{\alloc_1}, \dots, \overline{\alloc_{j-1}}$ given the (possibly randomized) algorithm $\alloc$.
\end{definition}

Notice that for uniformity, we discount by $\lambda^j$ \textit{all} transactions that are available in round $j$. However, for the purpose of the competitive analysis (which we soon detail), this is equivalent to the model where the adversary chooses an initial (un-discounted) weight for the packet, which is then discounted at every round the packet remains available. This is because the adversary can choose $w \cdot \lambda^{-j}$ for a new packet introduced at round $j$, and achieve the same effect as in our motivating setup.

\cref{def:TotalUtility} takes a nice form in the deterministic case for a given discount $\discount$:
\begin{equation}
\label{eq:detUtility}
        \revenue(\alloc | \adversary)
        =
        \sum_{j=0}^\horizon
        \discount^j \cdot \fee_{\alloc(\adversary(j)\cup \pending(j))}
    .
\end{equation}

To evaluate the performance of different allocation algorithms, we examine a game between an allocator and an adversary who creates transactions arrival designed to minimize the allocator's revenue.
The main challenge of this game is that the allocator has no foresight of future transactions that will be sent by the adversary.

The quality of an online allocation algorithm $\alloc$ when faced with a worst-case adversary $\adversary$ is quantified by dividing the utility $\alloc$ obtains with the utility obtained by the best possible offline algorithm $\alloc'$.
The resulting quantity is called $\alloc$'s \emph{competitive ratio}, and is formalized using our notations in \cref{def:CompRatio}.
A lower-bound is then attained by finding an allocation algorithm that guarantees good performance, and an upper-bound is attained by showing that no allocation algorithm can guarantee better performance.

\begin{definition}[Competitive Ratio]
    \label{def:CompRatio}
    Given an allocation algorithm $\alloc:(\txDomain)^* \rightarrow (\txDomain)^\blocksize$, its \emph{competitive ratio} under block-size $\blocksize$ is defined as:
    $
        R_{\alloc}
        \define
        \min_{\adversary} \frac{\revenue(\alloc | \adversary)}{\max_{\alloc'} \revenue(\alloc' | \adversary)}
    $.
\end{definition}
\begin{remark}
    For simplicity, \cref{def:CompRatio} uses $\max_{\alloc'} \revenue(\alloc' | \adversary)$ rather than $\sup_{\alloc'} \revenue(\alloc' | \adversary)$.
    This is in-line with the literature, such as the classic work of \cite{hajek2001competitiveness}
    .
    \cref{def:CompRatio}, moreover, implies that the competitive ratio of any algorithm $\alloc$ is $R_{\alloc} \leq 1$, with a ratio closer to $1$ implying that the performance of $\alloc$ is closer to optimal.
    We find this approach more natural, but many works adopt an inverse definition, so it is important to notice this choice.
\end{remark}

\section{The Deterministic Case}
\label{sec:Deterministic}
In this section, we focus on the deterministic case for our discounted model with some discount rate $\discount$.
Missing proofs are given in \cref{sec:Proofs}.

\label{sec:DeterministicUpperBound}

Previous competitive ratio bounds for the undiscounted case rely on constructions that do not capture the discounted case, and cannot apply there.
For example, the deterministic undiscounted upper bound equals one over the golden ratio $\frac{1}{\phi} \approx 0.618$~\cite{vesely2019a}, but as we see in \cref{res:GreedyCompRatio}, even the naïve $Greedy$ algorithm outperforms this bound for some discount factors $\discount$.
In \cref{res:DeterministicUpperBound}, we prove an upper bound on the performance of any allocation algorithm in this broader setting.

\begin{restatable}[Deterministic Upper Bound]{theorem}{resDeterministicUpperBound}
    \label{res:DeterministicUpperBound}
    Let $\imratio = \frac{1}{2}(\discount + \sqrt{\discount^2 + 4})$.
    Then, any deterministic allocation algorithm $ALG$ has $R_{ALG} \leq \frac{1}{\imratio}$.
\end{restatable}

\begin{proof}[Proof sketch]
We prove the bound for $\discount \imratio < 1$, which happens when $\discount < \frac{1}{\sqrt{2}} \approx 0.707107$.
The more complex proof for the general case appears in \cref{sec:Proofs}.

Consider a series of adversaries $\{\adversary_n\}_{n=1}^\infty$, defined inductively below.
\begin{align}
    \forall n \in \mathbb{N}:
    \adversary_n(i)
    &\define
    \begin{cases}
        \emptyset, & \substack{i\le0 \text{ or } n<i} \\
        \adversary_{n-1}(i), & \substack{i < n} \\
        \left\{
            \left(1, \imratio^{n-1}\right),
            \left(2, \imratio^n\right)
        \right\}, & \substack{i = n} \\
    \end{cases}, \\
    \adversary_{\infty}(i)
    &\define
    \adversary_i(i).
\end{align}

Notice that $\psi_{\infty}$ does not formally fit our framework, as we only consider finite schedules. However, since $\lambda \ell < 1$, for large enough $N$, the total effect of the tail of the schedule is negligible, and so we can consider $\psi_{N}$ that contains only the first $N$ terms of $\psi_{\infty}$ instead. 

We first illustrate the construction.
For $n=1$, we get that $\adversary_1(1) = \left\{ \left(1, 1\right), \left(2, \imratio\right) \right\}$, and $\forall i \ne 1: \adversary_1(i) = \emptyset$.
Thus, the adversary sends two transactions in step $1$, and does not broadcast any other transaction.
The first of these transactions, $\left(1, 1\right)$, pays a fee equal to $1$ and expires at the end of the turn.
The second transaction, $\left(2, \imratio\right)$, expires at the end of the next turn and pays a fee of $\imratio$.

The adversary $\adversary_2$ offers two transactions in step $2$: $\adversary_2(2) = \left\{ \left(1, \imratio\right), \left(2, \imratio^2\right) \right\}$.
Similarly to the previous case, the former transaction expires at the end of step $2$, and the latter at the end of step $3$.
The general case where $n > 1$ is depicted in \cref{fig:DeterministicUpperBoundTTL1}.

Consider any deterministic algorithm $ALG$, and let $n + 1$ be the first index where it allocates a transaction with $TTL=2$ when facing the adversary $\adversary_{\infty}$.
We now split the proof according to whether such an index exists or not.

\subsubsection*{Case I: No Such Index Exists}
Due to our choice of $\imratio = \frac{1}{2}(\discount + \sqrt{\discount^2 + 4})$, then $\imratio \geq 1$ for any $0 \leq \discount \leq 1$.
Thus, we compare $ALG$'s performance with $Greedy$ that allocates the highest-fee transaction at each step:
\begin{align*}
    R_{ALG}
    &
    =
    \min_{\adversary} \frac{\revenue(ALG | \adversary)}{\max_{\alloc'} \revenue(\alloc' | \adversary)}
    \nonumber
    \le
    \frac{\revenue(ALG | \adversary_{\infty})}{\max_{\alloc'} \revenue(\alloc' | \adversary_{\infty})}
    \nonumber
    \le
    \frac{\revenue(ALG | \adversary_{\infty})}{\revenue(Greedy | \adversary_{\infty})}
    \nonumber
    =
    \frac{\sum_{i=1}^{\infty} \discount^i\imratio^{i-1}}{\sum_{i=1}^{\infty} \discount^i\imratio^i}
    \nonumber
    =
    \frac{1}{\imratio}.
\end{align*}

\begin{figure}[t]
    \centering
    \scalebox{0.9}{\begin{tikzpicture}[
  LeftArr/.style={shape=circle, minimum size=0.18cm, inner sep=0pt, draw=white, fill=white, label=center:$\leftarrow$},
  RightArr/.style={shape=circle, minimum size=0.18cm, inner sep=0pt, draw=white, fill=white, label=center:$\rightarrow$},
  Greedy/.style={shape=rectangle, minimum size=0.2cm, inner sep=0pt, draw=mediumseagreen85168104, fill=mediumseagreen85168104},
  ALG/.style={shape=circle, minimum size=0.2cm, inner sep=0pt, draw=indianred1967882, fill=indianred1967882},
]
\tikzmath{
    \FirstStepNum   = 3;
    \LastStepNum    = 3;
    \StepNum        = \FirstStepNum + \LastStepNum + 1;
    \FirstTxNum     = 4;
    \LastTxNum      = 2;
    \TxNum          = \FirstTxNum + \LastTxNum;
    \MidTx          = -\TxNum/2-0.5;
}

% START Vertical lines
\node[] at (\StepNum/2+0.5, 0.5) {Step};
\foreach \x in {1, ..., \StepNum} {
    \draw[gray, dashed] (\x, -0.5) -- +(0, -\TxNum);
}
\foreach \x in {1, ..., {\FirstStepNum}} {
    \node[] at (\x+0.5, 0) {$\x$};
}
\node[] at (\FirstStepNum+1.5, 0) {...};
\node[] at (\FirstStepNum+2.5, 0) {$n$};
\node[] at (\FirstStepNum+3.5, 0) {$n+1$};
% END Vertical lines

% START TXs
\node[rotate=90] at (-2, \MidTx) {Transaction (\gls{TTL}, fee)};

\foreach \x in {1, 2, 3} {
    \tikzmath{
        \y = 1-2*\x;
        if \x == 3 then {
            \x = 5;
            let \idxA = n-1;
            let \idxB = n;
        } else {
            int \idxA;
            \idxA = \x-1;
            \idxB = \x;
        };
    };
    \draw[decorate, decoration={calligraphic brace, mirror}] (-0.5, {\y+0.25}) --  +(0, -1.5) node[left, midway]{$\adversary_n(\idxB)$};
    \node[] at (0.25, \y) {$(1, \imratio^{\idxA})$};
    \draw[<->] (\x, \y) -- +(1, 0);
    \node[] at (0.25, \y-1) {$(2, \imratio^{\idxB})$};
    \draw[<->] (\x, \y-1) -- +(2, 0);
}

\node[] at (0.25, -4.4) {$\vdots$};
% END TXs

% START Allocations
% Step 1
\node[ALG] at (1.5, -1) {};
\node[Greedy] at (1.5, -2) {};

% Step 2
\node[ALG] at (2.5, -3) {};
\node[Greedy] at (2.5, -4) {};

% Step n
\node[ALG] at (5.5, -5) {};
\node[Greedy] at (5.5, -6) {};

% Step n+1
\node[ALG] at (6.5, -6) {};

% \baselineskip is required to make sure the two lines won't overlap
% END Allocations

% START Legend
\matrix[draw, right] at (\StepNum+0.5, \MidTx) {
    \node[LeftArr, label=right:Broadcast] {}; \\
    \node[RightArr, label=right:Expiration] {}; \\
    \node[ALG, label=right:$ALG$] {}; \\
    \node[Greedy, label=right:$Greedy$] {}; \\
};
% END Legend
\end{tikzpicture}}
    \caption{
        A depiction of the first scenario presented in \cref{res:DeterministicUpperBound}, where the algorithm $ALG$ always allocates transactions with a \gls{TTL} of $1$ up until and including step $n$.
        In some cases, $ALG$ has multiple equivalent choices, which do not affect the analysis.
        For example, in the $n+1$ step $ALG$ can choose between two transactions that expire at the end of the turn and have a fee of $\imratio^1$: Either the one that carries over from step $n$, or the $(1, \imratio^n)$ transaction introduced in step $n+1$.
        Because $Greedy$ picks the highest-fee transactions, at step $n$ it allocates the transaction $(2, \imratio^n) \in \adversary_n(n)$.
    }
    \label{fig:DeterministicUpperBoundTTL1}
\end{figure}

\subsubsection*{Case II: the Index Exists}
Consider the case where $ALG$ eventually allocates a transaction with \gls{TTL} equal to $2$, meaning that $n+1$ is well defined.
Given our construction, this necessarily means that there will be no transactions to carry over from turn $n+1$ to $n+2$.

\begin{figure}
    \centering
    \scalebox{0.9}{\begin{tikzpicture}[
  LeftArr/.style={shape=circle, minimum size=0.18cm, inner sep=0pt, draw=white, fill=white, label=center:$\leftarrow$},
  RightArr/.style={shape=circle, minimum size=0.18cm, inner sep=0pt, draw=white, fill=white, label=center:$\rightarrow$},
  Greedy/.style={shape=rectangle, minimum size=0.2cm, inner sep=0pt, draw=mediumseagreen85168104, fill=mediumseagreen85168104},
  ALG/.style={shape=circle, minimum size=0.2cm, inner sep=0pt, draw=indianred1967882, fill=indianred1967882},
]
\tikzmath{
    \FirstStepNum   = 3;
    \LastStepNum    = 4;
    \StepNum        = \FirstStepNum + \LastStepNum + 1;
    \FirstTxNum     = 4;
    \LastTxNum      = 4;
    \TxNum          = \FirstTxNum + \LastTxNum;
    \MidTx          = -\TxNum/2-0.5;
}

% START Vertical lines
\node[] at (\StepNum/2+0.5, 0.5) {Step};
\foreach \x in {1, ..., \StepNum} {
    \draw[gray, dashed] (\x, -0.5) -- +(0, -\TxNum);
}
\foreach \x in {1, ..., {\FirstStepNum}} {
    \node[] at (\x+0.5, 0) {$\x$};
}
\node[] at (\FirstStepNum+1.5, 0) {...};
\node[] at (\FirstStepNum+2.5, 0) {$n$};
\node[] at (\FirstStepNum+3.5, 0) {$n+1$};
\node[] at (\FirstStepNum+4.5, 0) {$n+2$};
% END Vertical lines

% START TXs
\node[rotate=90] at (-3, \MidTx) {Transaction (\gls{TTL}, fee)};

\foreach \x in {1, 2, 3, 4} {
    \tikzmath{
        \y = 1-2*\x;
        if \x == 3 then {
            \x = 5;
            let \idxA = n-1;
            let \idxB = n;
        } else {
            if \x == 4 then {
                \x = 6;
                let \idxA = n;
                let \idxB = n+1;
            } else {
                int \idxA;
                \idxA = \x-1;
                \idxB = \x;
            };
        };
    };
    \draw[decorate, decoration={calligraphic brace, mirror}] (-0.5, {\y+0.25}) --  +(0, -1.5) node[left, midway]{$\adversary_{n+1}(\idxB)$};
    \node[] at (0.25, \y) {$(1, \imratio^{\idxA})$};
    \draw[<->] (\x, \y) -- +(1, 0);
    \node[] at (0.25, \y-1) {$(2, \imratio^{\idxB})$};
    \draw[<->] (\x, \y-1) -- +(2, 0);
}

\node[] at (0.25, -4.4) {$\vdots$};
% END TXs

% START Allocations
% Step 1
\node[ALG] at (1.5, -1) {};
\node[Greedy] at (1.5, -2) {};

% Step 2
\node[ALG] at (2.5, -3) {};
\node[Greedy] at (2.5, -4) {};

% Step n
\node[ALG] at (5.5, -5) {};
\node[Greedy] at (5.5, -6) {};

% Step n+1
\node[ALG] at (6.5, -8) {};
\node[Greedy] at (6.5, -7) {};

% Step n+2
\node[Greedy] at (7.5, -8) {};
% END Allocations

% START Legend
\matrix[draw, right] at (\StepNum+0.5, \MidTx) {
    \node[LeftArr, label=right:Broadcast] {}; \\
    \node[RightArr, label=right:Expiration] {}; \\
    \node[ALG, label=right:$ALG$] {}; \\
    \node[Greedy, label=right:$Greedy'$] {}; \\
};
% END Legend
\end{tikzpicture}}
    \caption{The setting described in the proof of the deterministic upper bound (see \cref{res:DeterministicUpperBound}), for the case where $ALG$ picks a transaction with a \gls{TTL} equal to $2$.}
    \label{fig:DeterministicUpperBoundTTL2}
\end{figure}

Given the adversary $\adversary_{n+1}$, depicted in \cref{fig:DeterministicUpperBoundTTL2}, we compare the performance of $ALG$ with the allocation $Greedy'$ that allocates greedily in all steps $1 \leq i \leq n$, then allocates at step $n+1$ a transaction with $TTL=1$, and finally picks the single remaining transaction at step $n+2$.
One can then verify that $ALG$ achieves the (weakly) best ratio against $Greedy'$ when $n=0$ (thus minimizing to zero the number of steps where $Greedy'$ allocates a higher-fee transaction greedily).
By our construction, the latter transaction pays a fee of $\imratio^{n+1}$, thus:
\begin{align}
    \label{eq:UpperBoundFirstStep}
    R_{ALG}
    &
    =
    \min_{\adversary} \frac{\revenue(ALG | \adversary)}{\max_{\alloc'} \revenue(\alloc' | \adversary)}
    \nonumber
    \leq
    \min_n \frac{\revenue(ALG | \adversary_{n+1})}{\revenue(Greedy' | \adversary_{n+1})}
    \nonumber
    \leq
    \frac{\imratio}{\revenue(Greedy' | \adversary_1)}
    \nonumber
    =
	\frac{\imratio}{1 + \discount \imratio}
    \nonumber
    =
    \frac{1}{\imratio}
    .
\end{align}
The last equation holds due to having $\imratio = \frac{1}{2}(\discount + \sqrt{\discount^2 + 4})$, which implies $2 \imratio - \discount = \sqrt{\discount^2 + 4}$, and thus also that $4 \imratio^2 - 4 \imratio \discount +\discount^2 = \discount^2 + 4$, allowing to conclude that $\imratio^2 = 1 + \discount \imratio$.
\end{proof}

\subsection{The Immediacy-Biased Class Of Allocation Algorithms}
\label{sec:ImmediacyBiased}
We proceed by introducing the \emph{immediacy-biased} class of allocation algorithms, and identify a regime of discount rates $\discount$ which we call the ``semi-myopic'' regime where it achieves the optimal deterministic competitive ratio.
Given a parameter $\imratio \in \mathcal{R}$, we denote the corresponding instance of this class as ${{\imratio}IB}$ and define it in the following manner:
\begin{definition}[The $\imratio$-immediacy-biased Algorithm (${{\imratio}IB})$]
\label{def:ImmediacyBiased}
Let $\fee_{max}, \fee_{urg}$ respectively be the highest-fee available transaction, and the lowest-\gls{TTL} transaction (i.e., a transaction with a \gls{TTL} of $1$) that offers the highest fee.
If $\frac{\fee_{max}}{\fee_{urg}} \geq \imratio$, allocate the transaction matching $\fee_{max}$; otherwise, allocate the one matching $\fee_{urg}$. 
\end{definition}

Intuitively, the $\imratio IB$ algorithm is somewhat biased towards greediness, but the best transaction that immediately expires has a preference over the best transaction that expires in future blocks, as long as it is not $\imratio$ times worse, or more.
In particular, we are interested in the case where $\imratio =  \frac{1}{2}(\discount + \sqrt{\discount^2 + 4})$.
This choice is justified since we show we can upper bound $\imratio IB$ allocation for any $\imratio$ parameter by both $\frac{1}{\imratio}$ and $\frac{\imratio}{1 + \discount \imratio}$ (\cref{res:GoldenLower}), so the only way to achieve the optimal upper bound of \cref{res:DeterministicUpperBound} is by making this choice.
Moreover, we are able to precisely pinpoint the regime for which it is optimal, which is whenever $\discount \lesssim 0.77$.
This is since we prove an upper bound of $\frac{1+\discount \imratio}{1 + \discount + \discount^2 \imratio + \discount^3}$, and a lower bound of $\frac{1}{1+\discount^3}$.
As \cref{fig:CompetitiveRatios} shows, these bounds take precedence over the $\frac{1}{\imratio}$ bound exactly at $\approx 0.77$, where
$$\frac{1+\discount \imratio}{1 + \discount + \discount^2 \imratio + \discount^3} = \frac{1}{1+\discount^3} = \frac{1}{\imratio}.$$

We bound the allocation algorithm's competitive ratio from below in \cref{res:GoldenGreedy}.
\begin{restatable}[]{theorem}{resGoldenGreedy}
    \label{res:GoldenGreedy}
    Let $\imratio = \frac{1}{2}(\discount + \sqrt{\discount^2 + 4})$, then:
    $R_{{\imratio}IB} \geq \min \left \{\frac{1}{\imratio}, \frac{1}{1 + \discount^3} \right \}$.
\end{restatable}

\begin{proof}[Proof Sketch.]
The proof has the following structure.
We perform a step-by-step comparison of transactions chosen by our algorithm (ALG) and the optimal allocation (OPT).
An important distinction is whether the transaction allocated by OPT at a certain step $i_0$ is also available to ALG at the same step.
If available, then the transaction allocated by ALG at that step must be at most $\ell$-times worse (as ALG always chooses a transaction that is at most $\ell$-times worse than the best transaction available to it). 

If the transaction is not available, it must be because ALG has already previously allocated it at some step $i_1$. We then check whether at step $i_1$ the transaction allocated by OPT was available to ALG. If not, we continue in this fashion, until we must reach some $i_k$ for which this condition is true. We then examine this \textit{chain} of steps $i_k, \ldots, i_0$ in isolation (notice that because of our backwards induction construction, the earlier steps have the higher indices). An important property of such a chain is that the transaction allocated by ALG at step $i_{j+1}$ is allocated by OPT at step $i_j$, and so it yields a lower (discounted) value. 

Consider the following two cases: 

\begin{itemize}
    \item If the TTL of the transaction allocated by OPT at step $i_k$ is $1$, then since it also available to ALG (as this step terminates the backward induction chain construction), it is a lower bound to $\fee_{urg}$. However, we know that ALG allocates $\fee_{max}$, because we have a chain, which means OPT allocates ALG's transaction at a later step (and so it is not possible that ALG allocates an urgent transaction). We conclude that at step $i_k$, ALG allocates a transaction with value \textit{at least} $\ell$ times that of OPT's allocated transaction at step $i_k$. 

    For all other transactions allocated by OPT, we note that an important property of \textit{any} chain is that the transaction allocated by ALG at step $i_{j+1}$ is allocated by OPT at $i_j$, and so it yields a lower, discounted value. Overall, this guarantees that along the chain, OPT generates at most $\frac{1 + \imratio \discount}{\imratio}$ more value than ALG. For our choice of $\imratio$, this equals exactly $\frac{1}{\imratio}$. 

    \item If $i_{k-1}$ is $3$ steps or more after $i_k$, then, first note that ALG has at least the same value as OPT at $i_k$ (since it has OPT's allocated transaction available, and it allocates the maximal available transaction, as previously shown).
    Moreover, when OPT allocates this transaction at $i_{k-1}$, it is discounted by at least $\discount^3$.
    Together with the fact that OPT always chooses transactions later than ALG along the chain, this yields a $(1 + \lambda^3)$ ratio.
\end{itemize}

Thus, the remaining cases are those where $i_{k-1}$ either immediately follows $i_k$, or is two steps away.
This allows for a detailed case analysis, since we only need to account for the transactions allocated in between $i_k$ and $i_{k-1}$, which does not introduce too many variables.
We show that these cases do not introduce worse ratios in our full proof (see \cref{sec:Proofs}). 
\end{proof}

In \cref{sec:Proofs}, we also show an upper bound for the $\imratio IB$ algorithm analysis.
\begin{restatable}{proposition}{resGoldenUpper}
\label{res:GoldenUpper}
Let $\imratio = \frac{1}{2}(\discount + \sqrt{\discount^2 + 4})$, then:
$$
    R_{{\imratio}IB} \leq \min \left \{\frac{1}{\imratio}, \frac{1+\discount \imratio}{1 + \discount + \discount^2 \imratio + \discount^3},\min_{n\geq 1} \frac{\sum_{i=0}^{n+1} \discount^i}{\sum_{i=0}^{2n} \discount^i} \right \}
    .
$$
\end{restatable}

\section{The Randomized Case}
\label{sec:Randomized}

In this section, we consider randomized allocation algorithms.
Thus, we should clarify how the adversary may react, as the previous setting only accounted for fixed deterministic adversaries.
We consider the stronger \emph{adaptive} adversary, that may randomize its own transaction arrivals, and also has the information of past choices of the algorithm and can use it when choosing its future course of action.

\begin{restatable}[Randomized Upper Bound]{theorem}{resRandomizedUpperBound}
\label{res:RandomizedUpperBound}
    For any (possibly randomized) allocation algorithm $ALG$ we have that the competitive ratio has:
    $R_{ALG} \leq 1 - \frac{\discount}{4}.$
\end{restatable}
\begin{proof}[Proof Sketch.]
Our proof follows along the lines of the proof for the undiscounted case given in \cite{bienkowski2011randomized}.
We use a recursive construction of adversaries where the fees grow exponentially by a factor of $b = \frac{2}{2-\discount}$.
The choice of the exponent's base $b$ is crucial for the analysis.
To provide intuition for this choice, denote the probability (that in full generality may be step-specific) that at some step the algorithm allocates an immediate transaction with $TTL = 1$ and value $1$, rather than a transaction with $TTL = 2$ and value $b$. Then, the algorithm's utility for this step (disregarding later effects) is $p + (1-p) \cdot b$. The reason we can disregard later effects is because we can have the adaptive adversary act in the following way: If the algorithm allocates the immediate transaction, the adversary adds transactions in the next round that make ``delaying gratification'' unnecessary (you could have chosen the transaction with value $b$, and still get high-value transactions later). If the algorithm allocates the high-value transaction, then the adversary terminates the interaction (does not offer any more transactions). We then consider the algorithm's performance in comparison with ADV, which is an alternative algorithm we bundle together with the adversary's choice for the two outcomes: (i) The adversary terminates, and ADV allocates $1$ at the current step and $b$ at the next. (ii) The adversary adds a high-value transaction, in which case ADV allocates $b$ at the current step, and also (with probability $p$, which is the probability the algorithm ALG allocates the low-value transaction in the current step) allocates a high-value transaction in the next step. 
If we divide ALG's performance by ADV's performance in the two cases, we get the following equation, which turns out to work well for the choice of $b$:
$$\frac{p + (1-p) \cdot b}{1 + \discount \cdot b} = \frac{p + (1-p) \cdot b}{b + p\cdot \discount \cdot b}.$$
In particular, we use its solution for $p=\frac{1}{2}$, which is the ``most balanced'' case. 
\end{proof}

\subsection{The \texorpdfstring{$RDISC$}{RDISC} Randomized Allocation Algorithm}
\label{sec:RDISC}
We now devise a randomized allocation algorithm and prove that its performance guarantees exceed those attained by the best possible deterministic algorithm.
Our construction takes some inspiration from the classic $RMIX$ algorithm of \cite{chin2006online} for the undiscounted case, which, in turn, builds on the preceding work of \cite{chrobak2003preemptive}.

To fit the more general discounted setting, we enrich our algorithm with a parametrization to account for the rate $\discount$.
As the proof shows, this parametrization is delicate and cannot be done naïvely: the sampling step that $RDISC$ relies on should be carefully performed when considering the economy's discount factor.
\begin{definition}[$RDISC_{\discount}$]
\label{def:RDISC}
Let $\fee_{max}, \fee_{urg}$ respectively be the highest-fee available transaction, and the lowest-\gls{TTL} transaction that offers the highest fee.
$RDISC_{\discount}$ is the randomized algorithm that samples $\theta \sim UNI([-\discount, 0])$ (the uniform distribution over the interval $[-\discount, 0]$), and allocates $\fee_{urg}$ if and only if $\fee_{urg} \geq e^{\theta} \cdot \fee_{max}$, and otherwise allocates $\fee_{max}$.
\end{definition}

Compared to RDISC, RMIX under-prioritizes high-value transactions for $\discount < 1$.
Consider transactions $(1, e + 1), (2, e^2)$ with $\discount = \frac{1}{e}$. RMIX has a positive probability of choosing the earlier deadline packet first, obtaining a discounted value of $2e + 1$.
On the other hand, our $RDISC_{\frac{1}{e}}$ always chooses $(2, e^2)$ first, and thus obtains a value of $e^2$.

\begin{restatable}[]{theorem}{resRDISC}
\label{res:RDISC}
    $RDISC_{\discount}$ has $R_{RDISC_{\discount}} \geq \frac{1 - e^{-\discount}}{\discount}$.
\end{restatable}

\begin{proof}
We show the proof for an oblivious adversary using a potential function argument. \cite{bienkowski2011randomized} show how to extend this argument (in the undiscounted case) to an adaptive adversary. 

If $\Phi$ is a potential function that receives a set of transactions, returns $0$ on the empty set, and satisfies for any step $i$:
$R \fee_{ALG}(\adversary(i) \cup \pending(i)) \geq  \fee_{OPT}(\adversary(i) \cup \pending(i)) + \discount \Phi(\adversary(i+1) \cup \pending(i+1)) - \Phi(\adversary(i) \cup \pending(i)),$
then $ALG$ has a competitive ratio of $\frac{1}{R}$.
Proof by summation:
\[
\begin{split}
    &
    R\sum_{i=0}^N  \discount^i \fee_{ALG}(\adversary(i) \cup \pending(i))
    \\&
    \geq
    \sum_{i=0}^N \left( \discount^i \fee_{OPT}(\adversary(i) \cup \pending(i)) + \discount\left(\Phi( \adversary(i+1) \cup \pending(i+1)) - \Phi(\adversary(i) \cup \pending(i)) \right) \right)
    \\&
    =
    \sum_{i=0}^N  \discount^i \fee_{OPT}(\adversary(i) \cup \pending(i))
    %\\&
    +
    \sum_{i=0}^N  \discount^i \left(\discount  \Phi(\adversary(i+1) \cup \pending(i+1)) - \discount \Phi(\adversary(i) \cup \pending(i)) \right)
\end{split}
\]
It remains to show that:
$
\sum_{i=0}^N \discount^i \left(\discount \Phi(\adversary(i+1) \cup \pending(i+1)) -\Phi(\adversary(i) \cup \pending(i)) \right)
\geq
0
.
$
Notice that $\Phi(\adversary(0) \cup \pending(0)) = 0$, and so:
\[
\begin{split}
    &
    \discount \Phi(\adversary(1) \cup \pending(1)) + \sum_{i=1}^N  \discount^{i+1} \Phi(\adversary(i+1) \cup \pending(i+1)) - \sum_{i=1}^N  \discount^i \Phi(\adversary(i) \cup \pending(i))
    \\&
    \geq
    \sum_{i=1}^N  \discount^{i+1} \Phi(\adversary(i+1) \cup \pending(i+1)) - \sum_{i=2}^N \discount^i\Phi(\adversary(i) \cup \pending(i))
    \\&
    =
    \sum_{i=2}^{N+1} \discount^{i} \Phi(\adversary(i) \cup \pending(i)) - \sum_{i=2}^N  \discount^{i} \Phi(\adversary(i) \cup \pending(i))
    \\&
    =
     \discount^{N+1} \Phi(\adversary(N+1) \cup \pending(N+1))
    % \\&
    =
    0
    .
\end{split}
\]
It remains to show that at every step $i$, there is such a potential function $\Phi$ with:

$$\fee_{RDISC_{\discount}}(\adversary(i) \cup \pending(i)) \geq \frac{1-e^{-\discount}}{\discount} \fee_{OPT}(\adversary(i) \cup \pending(i)) + \discount\Phi(\adversary(i+1) \cup \pending(i+1)) - \Phi(\adversary(i) \cup \pending(i)).$$

Let $\Gamma(\discount) = \frac{\discount}{1 - e^{-\discount}}$.
We define $\fee_{RDISC_{\discount}} = \fee_{RDISC_{\discount}}(\adversary(i) \cup \pending(i))$, and $\fee_{OPT} = \fee_{OPT}(\adversary(i) \cup \pending(i))$.
The proof of Theorem 3.1 in \cite{chin2006online} shows that there is a potential function where for any $\theta \in \left[-1, \log{\frac{E[\fee_{OPT}]}{w_h}}\right]$, $\fee_{RDISC_{\discount}} - (\Phi(\adversary(i+1) \cup \pending(i+1)) - \Phi(\adversary(i) \cup \pending(i))) = \fee_{RDISC_{\discount}}$, and for any $\theta \in \left[\log{\frac{E[\fee_{OPT}]}{w_h}}, 0\right]$, $\fee_{RDISC_{\discount}} - (\Phi(\adversary(i+1) \cup \pending(i+1)) - \Phi(\adversary(i) \cup \pending(i))) \geq 0$. 
Let $w_h$ be the highest-fee packet available to $RDISC_{\discount}$ at $i$.
Thus, by sampling uniformly from $[-\lambda, 0]$:

\begin{equation}
\label{eq:chin_et_al}
\begin{split}
& E\left[\fee_{RDISC_{\discount}} - (\Phi(\adversary(i+1) \cup \pending(i+1)) - \Phi(\adversary(i) \cup \pending(i)))\right] \\
& = \int_{-\lambda}^0 \frac{1}{\lambda}\left(\fee_{RDISC_{\discount}} - (\Phi(\adversary(i+1) \cup \pending(i+1)) - \Phi(\adversary(i) \cup \pending(i)))\right) \\
& = \int_{-\lambda}^{\log{\frac{E[\fee_{OPT}]}{w_h}}} \frac{1}{\lambda}\left(\fee_{RDISC_{\discount}} - (\Phi(\adversary(i+1) \cup \pending(i+1)) - \Phi(\adversary(i) \cup \pending(i)))\right) \\
& + \int_{\log{\frac{E[\fee_{OPT}]}{w_h}}}^{0} \frac{1}{\lambda}\left(\fee_{RDISC_{\discount}} - (\Phi(\adversary(i+1) \cup \pending(i+1)) - \Phi(\adversary(i) \cup \pending(i)))\right) \\
& \geq \int_{-\discount}^{\log{\frac{E[\fee_{OPT}]}{w_h}}} \frac{1}{\discount} e^{\theta} w_h d\theta + 0,
\end{split}
\end{equation}

With this potential function, we can thus write at step $i$,
\[
\begin{split}
    &
    E\left[\Gamma(\discount) \fee_{RDISC_{\discount}} - (\discount \Phi(\adversary(i+1) \cup \pending(i+1)) - \Phi(\adversary(i) \cup \pending(i)))\right]
    \\&
    =
    ( \Gamma(\discount) - \discount) \cdot E\left[\fee_{RDISC_{\discount}}\right] + \discount E\left[\fee_{RDISC_{\discount}} - (\Phi(\adversary(i+1) \cup \pending(i+1)) - \frac{1}{\discount}\Phi(\adversary(i) \cup \pending(i)))\right]
    \\&
    \geq
    ( \Gamma(\discount) - \discount) \cdot E\left[\fee_{RDISC_{\discount}}\right] + \discount E\left[\fee_{RDISC_{\discount}} - (\Phi(\adversary(i+1) \cup \pending(i+1)) - \Phi(\adversary(i) \cup \pending(i)))\right]
    \\&
    \geq
    ( \Gamma(\discount) - \discount) \int_{-\discount}^0 \frac{1}{\discount} e^{\theta} w_h d\theta + \discount \int_{-\discount}^{\log{\frac{E[\fee_{OPT}]}{w_h}}} \frac{1}{\discount} e^{\theta} w_h d\theta
    \\&
    =
    w_h \cdot \left((\frac{1}{1-e^{-\discount}} - 1) (1 - e^{-\discount}) + \frac{E\left[\fee_{OPT}\right]}{w_h} - e^{-\discount}\right)
    =
    E\left[\fee_{OPT}\right]
    .
\end{split}
\]
\end{proof}

\section{Discussion}
\label{sec:Discussion}

\subsubsection*{Memorylessness}
We show that a novel memoryless algorithm achieves the deterministic upper bound in the semi-myopic case.
We provide a tight characterization of this optimality, and so the algorithm cannot achieve optimality with larger values of $\discount$.
When considering deterministic algorithms which are \emph{not} memoryless, \cite{vesely2019a} present an optimal construction for the undiscounted case.
An interesting direction for future work is to see if this elegant algorithm maintains its optimality in the discounted case (i.e., when $\discount < 1$).

\subsubsection*{General Time-dependent Models}
While multiplicative discounting is a natural model and common in the economic literature, other models can be of interest.
For example, one may consider packet-specific discounting, which can be taken to represent different rates of quality deterioration or inventory spoilage.

\subsubsection*{Semi-myopic Discounts}
The semi-myopic regime is relevant to notable settings such as long-term project management, where one has to decide which projects to commit resources to on a time scale of several years.
Moreover, there are alternative interpretations to the discount factor that show the usefulness of the semi-myopic range: One example is where clients, on top of having a strict deadline after which they leave the system, may also become impatient and leave following a geometric (memoryless) process.
Another example is the one we discussed regarding heterogeneous allocators: $\discount$ can be interpreted as a rule-of-thumb discount that stems from the probability of having allocators with different allocation rules decide the next block.
Intuitively, our analysis shows that a particular benefit of semi-myopic discount rates is that they lessen the impact of longer pathological chains of adversarial sequences.
We thus believe that extending our framework and using discounting could yield interesting results when applied to other online problems, for example, settings of online fairness such as the food bank problem \cite{aleksandrov2015online}. 

\section{Acknowledgements}
The AI-powered feedback tool Refine.ink was helpful in revising the consistency and clarity of the proofs in the paper. 

\bibliographystyle{splncs04}
\bibliography{main.bib}

\begin{thebibliography}{10}
\providecommand{\url}[1]{\texttt{#1}}
\providecommand{\urlprefix}{URL }
\providecommand{\doi}[1]{https://doi.org/#1}

\bibitem{aleksandrov2015online}
Aleksandrov, M., Aziz, H., Gaspers, S., Walsh, T.: Online fair division: Analysing a food bank problem (2015)

\bibitem{ashlagi2019edge}
Ashlagi, I., Burq, M., Dutta, C., Jaillet, P., Saberi, A., Sholley, C.: Edge {Weighted} {Online} {Windowed} {Matching}. In: Proceedings of the 2019 {ACM} {Conference} on {Economics} and {Computation}. pp. 729--742. {EC} '19, ACM, NY, NY, USA (6 2019). \doi{10.1145/3328526.3329573}

\bibitem{bienkowski2011randomized}
Bienkowski, M., Chrobak, M., Łukasz Jeż: Randomized competitive algorithms for online buffer management in the adaptive adversary model. Theoretical Computer Science  \textbf{412}(39),  5121--5131 (2011). \doi{10.1016/j.tcs.2011.05.015}

\bibitem{borodin2005online}
Borodin, A., El-Yaniv, R.: Online computation and competitive analysis. Cambridge University Press, Cambridge, United Kingdom (2005)

\bibitem{chin2006online}
Chin, F.Y., Chrobak, M., Fung, S.P., Jawor, W., Sgall, J., Tichý, T.: Online competitive algorithms for maximizing weighted throughput of unit jobs. Journal of Discrete Algorithms  \textbf{4}(2),  255--276 (2006). \doi{10.1016/j.jda.2005.03.005}

\bibitem{chrobak2003preemptive}
Chrobak, M., Epstein, L., Noga, J., Sgall, J., van Stee, R., Tichy, T., Vakhania, N.: Preemptive scheduling in overloaded systems. Journal of Computer and System Sciences  \textbf{67}(1),  183--197 (Aug 2003). \doi{10.1016/S0022-0000(03)00070-9}

\bibitem{dickerson2021allocation}
Dickerson, J.P., Sankararaman, K.A., Srinivasan, A., Xu, P.: Allocation problems in ride-sharing platforms: Online matching with offline reusable resources. ACM Transactions on Economics and Computation (TEAC)  \textbf{9}(3),  1--17 (2021). \doi{10.1145/3456756}

\bibitem{fisher1911purchasing}
Fisher, I.: The purchasing power of money: its' determination and relation to credit interest and crises. The Macmillan Company, New York, NY (1911), \url{https://fraser.stlouisfed.org/title/purchasing-power-money-3610}

\bibitem{fisher1930theory}
Fisher, I.: The Theory of Interest. The Macmillan Company, New York, NY (1930), \url{https://oll.libertyfund.org/titles/fisher-the-theory-of-interest}

\bibitem{hajek2001competitiveness}
Hajek, B.: On the competitiveness of on-line scheduling of unit-length packets with hard deadlines in slotted time. In: Proceedings of the 2001 Conference on Information Sciences and Systems. pp. 434--438. Johns Hopkins University, Baltimore, MD (2001), \url{https://hajek.ece.illinois.edu/Papers/Hajek01_OL.pdf}

\bibitem{kesselman2001buffer}
Kesselman, A., Lotker, Z., Mansour, Y., Patt-Shamir, B., Schieber, B., Sviridenko, M.: Buffer overflow management in {QoS} switches. In: Proceedings of the thirty-third annual {ACM} symposium on {Theory} of computing. pp. 520--529. {STOC} '01, Association for Computing Machinery, New York, NY, USA (Jul 2001). \doi{10.1145/380752.380847}

\bibitem{li2005optimal}
Li, F., Sethuraman, J., Stein, C.: An optimal online algorithm for packet scheduling with agreeable deadlines. In: Proceedings of the sixteenth annual {ACM}-{SIAM} symposium on {Discrete} algorithms. pp. 801--802. {SODA} '05, Society for Industrial and Applied Mathematics, USA (Jan 2005). \doi{10.5555/1070432.1070544}

\bibitem{liang2024learning}
Liang, Y.C., Stein, C., Wei, H.T.: Learning-{Augmented} {Online} {Packet} {Scheduling} with {Deadlines} (2024). \doi{10.48550/arXiv.2305.07164}

\bibitem{nahmias1982}
Nahmias, S.: Perishable inventory theory: A review. Operations Research  \textbf{30}(4),  680--708 (1982), \url{http://www.jstor.org/stable/170438}

\bibitem{tsanikidis2023optimal}
Tsanikidis, C., Ghaderi, J.: Near-{Optimal} {Packet} {Scheduling} in {Multihop} {Networks} with {End}-to-{End} {Deadline} {Constraints}. Proc. ACM Meas. Anal. Comput. Syst.  \textbf{7}(3),  50:1--50:32 (Dec 2023). \doi{10.1145/3626781}

\bibitem{van1963inventory}
Van~Zyl, G.J.: Inventory control for perishable commodities. Tech. rep., North Carolina State University. Dept. of Statistics (1963)

\bibitem{vesely2018online}
Vesel{\`y}, P.: Online Algorithms for Packet Scheduling. Ph.D. thesis, Univerzita Karlova, Matematicko-fyzik{\'a}ln{\'\i} fakulta (2018), \url{http://hdl.handle.net/20.500.11956/104429}

\bibitem{vesely2021packet}
Vesel\'{y}, P.: Packet scheduling: Plans, monotonicity, and the golden ratio. SIGACT News  \textbf{52}(2),  72–84 (Jun 2021). \doi{10.1145/3471469.3471481}

\bibitem{vesely2019a}
Vesel{\'{y}}, P., Chrobak, M., Jez, L., Sgall, J.: A {\(\phi\)}-competitive algorithm for scheduling packets with deadlines. In: Chan, T.M. (ed.) Proceedings of the Thirtieth Annual {ACM-SIAM} Symposium on Discrete Algorithms, {SODA} 2019, San Diego, California, USA, January 6-9, 2019. pp. 123--142. {SIAM}, Philadelphia, PA (2019). \doi{10.1137/1.9781611975482.9}

\end{thebibliography}

\appendix

\section{Revisiting the Greedy Allocation}
\label{sec:Greedy}
The $Greedy$ allocation (see~\cref{def:Greedy}) is a classic packet scheduling algorithm which was studied by prior work for the undiscounted case \cite{hajek2001competitiveness,kesselman2001buffer}.
As a warm-up for readers interested in familiarizing themselves with our discounted framework, we evaluate $Greedy$'s performance.

\begin{definition}[The $Greedy$ allocation algorithm]
    \label{def:Greedy}
    The $Greedy$ allocation algorithm allocates the highest paying transaction present in the set $S$ of all the currently available transactions, disregarding \gls{TTL}:
    \begin{equation*}
        Greedy(S) \define \argmax_{(\ttl, \fee) \in S} \fee
    \end{equation*}
    If there are multiple transactions with the same fee, these with the lowest \gls{TTL} are preferred.
\end{definition}

In \cref{exmp:GreedyDiscount}, we show how $Greedy$'s performance may depend on the discount rate.
\begin{example}
\label{exmp:GreedyDiscount}
    We examine $Greedy$'s performance given the following adversary $\adversary$.
    $$
        \adversary(i)
        \define
        \begin{cases}
            \{(1,2), (2,4)\}, & \substack{i=1} \\
            \{(2,6)\}, & \substack{i=2} \\
            \{(1,8)\}, & \substack{i=4} \\
            \emptyset, & \substack{\text{otherwise}} \\
        \end{cases}
    $$

    \begin{figure}
        \centering
        \scalebox{1.0}{\tikzstyle{scorestars}=[star, star points=5, star point ratio=2.25, draw, inner sep=1.3pt, anchor=outer point 3]
\begin{tikzpicture}[
  LeftArr/.style={shape=circle, minimum size=0.18cm, inner sep=0pt, draw=white, fill=white, label=center:$\leftarrow$},
  RightArr/.style={shape=circle, minimum size=0.18cm, inner sep=0pt, draw=white, fill=white, label=center:$\rightarrow$},
  Greedy/.style={shape=rectangle, minimum size=0.2cm, inner sep=0pt, draw=mediumseagreen85168104, fill=mediumseagreen85168104},
  OPT/.style={shape=star, star points=5, star point ratio=2.25, draw, inner sep=1.3pt, minimum size=0.18cm, draw=indianred1967882, fill=indianred1967882},
  OPTMyopic/.style={shape=circle, inner sep=1.3pt, minimum size=0.2cm, draw=steelblue76114176, fill=steelblue76114176},
]

\tikzmath{
    \FirstStepNum   = 4;
    \LastStepNum    = 0;
    \StepNum        = \FirstStepNum + \LastStepNum + 1;
    \FirstTxNum     = 4;
    \LastTxNum      = 0;
    \TxNum          = \FirstTxNum + \LastTxNum;
    \MidTx          = -\TxNum/2-0.5;
}

% START Vertical lines
\node[] at (\StepNum/2+0.5, 0.5) {Step};
\foreach \x in {1, ..., \StepNum} {
    \draw[gray, dashed] (\x, -0.5) -- +(0, -\TxNum);
}
\foreach \x in {1, ..., {\FirstStepNum}} {
    \node[] at (\x+0.5, 0) {$\x$};
}
% END Vertical lines

% START TXs
\node[rotate=90] at (-2, \MidTx) {Transaction (\gls{TTL}, fee)};

\draw[decorate, decoration={calligraphic brace, mirror}] (-0.5, {-1+0.25}) --  +(0, -1.5) node[left, midway]{$\adversary(1)$};
\node[] at (0.25, -1) {$(1, 2)$};
\draw[<->] (1, -1) -- +(1, 0);
\node[] at (0.25, -2) {$(2, 4)$};
\draw[<->] (1, -2) -- +(2, 0);

\draw[decorate, decoration={calligraphic brace, mirror}] (-0.5, {-3+0.25}) --  +(0, -0.5) node[left, midway]{$\adversary(2)$};
\node[] at (0.25, -3) {$(2, 6)$};
\draw[<->] (2, -3) -- +(2, 0);

\draw[decorate, decoration={calligraphic brace, mirror}] (-0.5, {-4+0.25}) --  +(0, -0.5) node[left, midway]{$\adversary(4)$};
\node[] at (0.25, -4) {$(1, 8)$};
\draw[<->] (4, -4) -- +(1, 0);
% END TXs

% START Allocations
% Step 1
\node[OPT] at (1.5, -1) {};
\node[Greedy] at (1.3, -2) {};
\node[OPTMyopic] at (1.7, -2) {};

% Step 2
\node[OPT] at (2.5, -2) {};
\node[Greedy] at (2.3, -3) {};
\node[OPTMyopic] at (2.7, -3) {};

% Step 3
\node[OPT] at (3.5, -3) {};

% Step 4
\node[Greedy] at (4.25, -4) {};
\node[OPT] at (4.75, -4) {};
\node[OPTMyopic] at (4.5, -4) {};
% END Allocations

% START Legend
\matrix[draw, right] at (\StepNum+0.5, \MidTx) {
    \node[LeftArr, label=right:Broadcast] {}; \\
    \node[RightArr, label=right:Expiration] {}; \\
    \node[OPT, label=right:\text{$OPT$ for $\discount = 1$}] {}; \\
    \node[OPTMyopic, label=right:\text{$OPT$ for $\discount = \frac{1}{4}$}] {}; \\
    \node[Greedy, label=right:$Greedy$] {}; \\
};
% END Legend
\end{tikzpicture}}
        \caption{
            An illustration of \cref{exmp:GreedyDiscount}.
            At every step $i$, the adversary broadcasts the transaction set $\adversary(i)$.
            Each transaction is denoted by a tuple of its \gls{TTL} and fee, and is depicted visually by a line stretching between the turn at which it was broadcast, and the one at which it expires (and becomes ineligible for inclusion).
            The algorithm $Greedy$ allocates the highest-fee transaction at each step, while $OPT$ is the optimal allocation given some discount rate $\discount$.
            The choices made at each step by $OPT$ for a discount rate of $\discount = 1$, $OPT$ for $\discount = \frac{1}{4}$ and $Greedy$ are denoted by a red star, blue circle, and a green rectangle, correspondingly.
            In the undiscounted case, meaning $\discount = 1$, the optimal utility is $20$ while $Greedy$ obtains a lower utility of $18$.
            For $\discount = \frac{1}{4}$, the allocation made by $Greedy$ coincides with the optimal one, and achieves a utility of $\frac{37}{32}$.
        }     
        \label{fig:GreedyDiscount}
    \end{figure}

    The adversarial arrival defined by $\adversary$ is depicted in \cref{fig:GreedyDiscount}.
    At turn $1$ the adversary broadcasts two transactions: $(1,2)$ which expires at the end of the turn and has a fee of $2$, and $(2,4)$ which pays a fee equal to $4$ and expires at the end of the next turn.
    Because $Greedy$ prioritizes transactions with higher fees, it will allocate $(2,4)$, while letting the other transaction expire.
    In the next turn, the adversary broadcasts a single transaction with a \gls{TTL} of $2$ and a fee of $6$, which is the only one available to $Greedy$ at that turn, and thus will be allocated.
    At step $3$, the adversary does not emit any transactions, and on step $4$, a transaction $(1,8)$ is broadcast and then allocated by $Greedy$.

    Assume a discount rate of $\discount = 1$.
    Given these parameters, $Greedy$'s allocation obtains a discounted utility of $18$ when facing $\adversary$.
    On the other hand, under the same setting, the optimal allocation, denoted by $OPT$, ``saves'' the transaction $(2,4) \in \adversary(1)$ to the second turn, and allocates $(1,2)$ in the first.
    Similarly, it waits until the third step to allocate $(2,6) \in \adversary(2)$, and in the final step it allocates $(1,8) \in \adversary(4)$.
    The total discounted utility for $OPT$ is $20$.

    Thus, given $\discount = 0.25$, the discounted utility of $Greedy$ equals $\frac{45}{32} = 4^{-1}\cdot4 + 4^{-2}\cdot6 + 4^{-4}\cdot8$.
    On the other hand, $OPT$ is heavily impacted by the choice of parameters.
    For this discount rate, the optimal allocation coincides with that of $Greedy$, while the previous allocation made by $OPT$ obtains a utility of $\frac{7}{8} = 4^{-1}\cdot2 + 4^{-2}\cdot4 + 4^{-3}\cdot6 + 4^{-4}\cdot8$.
    Thus, it is perhaps natural to expect that $Greedy$'s competitiveness improves with lower discount factors.
\end{example}

In \cref{res:GreedyCompRatio}, we derive the competitive ratio of $Greedy$ as a function of $\discount$. The proof represents a simple case to analyze using our subchains method, and is thus illustrative of the use of the method, which gets more complicated with our more advanced algorithms. We thus include it here in full. 

\begin{restatable}[]{proposition}{resGreedyCompRatio}
    \label{res:GreedyCompRatio}
    The competitive ratio of the $Greedy$ algorithm is $R_{Greedy} = \frac{1}{1+\discount}$.
\end{restatable}

\begin{proof}
We prove this result by separately showing the algorithm's lower and upper bounds.

\paragraph{Part I: $R_{Greedy} \leq \frac{1}{1+\discount}$}
Consider an adversary $\adversary$ with $\adversary(1) = \{(1,1), (2,1 + \epsilon)\}$ and $\adversary(i) = \emptyset$ for $i = 0$ and any $i \geq 2$.
Note that $\max_x \revenue(\alloc | \adversary) \geq \lambda (1 + \discount + \discount \epsilon)$.
By substituting these into \cref{def:TotalUtility} one can obtain the following utility:
$
    \revenue(Greedy | \adversary) = \lambda (1+\epsilon)
$.

\paragraph{Part II: $R_{Greedy} \geq \frac{1}{1+\discount}$}
Fix an adversary $\adversary$ and compare $Greedy$ with some algorithm $\alloc'$.

If $Greedy$ under-performs $\alloc'$ at index $i_0$,\footnote{When we write ``ALG1 under-performs ALG2 \emph{at}'' a certain index, we compare the algorithms' performances specifically at that point in time, \emph{not} cumulatively up to and including that point.} then it must be that the transaction chosen by $\alloc'$ is not available to $Greedy$ at this index.
As this transaction is available to $\alloc'$, it has not yet expired and thus must have been chosen by $Greedy$ at some prior index $i_1$.
If at $i_1$ $Greedy$ under-performs $\alloc'$, again it means that $\alloc'$ allocates a transaction that is not available to $Greedy$ and so must have been chosen by $Greedy$ at a previous index $i_2$, and so on.
Since $i_0$ is some finite index, there must be some number $k$ so that at $i_k$, $\alloc'$ allocates a transaction that is available to $Greedy$, and $Greedy$ weakly outperforms it.
Denote this ``chain'' of indices $C = \{i_k, \ldots, i_0\}$ (notice that by our choice of sub-indexing, $i_{j+1} < i_j$ for any $0 \leq j \leq k-1$).

First, note that since for $0 \leq j \leq k - 1$ the allocation of $Greedy$ at index $i_{j+1}$ is the same as the allocation of $\alloc'$ over the index $i_j$, we have:
\begin{equation}
\label{eq:GreedyChain}
\begin{split}
    \sum_{0 \leq j \leq k-1} \discount^{i_j}\fee(\alloc'(\adversary(i_j)\cup \pending_{\alloc'}(i_j)))
    &
    =
    \sum_{0 \leq j \leq k-1} \discount^{i_j}\fee(Greedy(\adversary(i_{j+1})\cup \pending_{Greedy}(i_{j+1})))
    \\&
    =
    \sum_{1 \leq j \leq k} \discount^{i_{j-1}}\fee(Greedy(\adversary(i_j)\cup \pending_{Greedy}(i_j)))
    \\&
    \leq \discount \sum_{1 \leq j \leq k} \discount^{i_j}\fee(Greedy(\adversary(i_j)\cup \pending_{Greedy}(i_j)))
\end{split}
\end{equation}

Notice our use of the fact that $i_{j-1}$ must be at least one round later than $i_j$, and so must be discounted by at least $\discount$ relative to it.

Since $Greedy$, by its definition, weakly outperforms $x'$ at $i_k$, we have:
\begin{equation}
    \label{eq:GreedyChainStart}
    \fee(\alloc'(\adversary(i_k)\cup \pending_{\alloc'}(i_k)))
    \leq
    \fee(Greedy(\adversary(i_k)\cup \pending_{Greedy}(i_k)))
\end{equation}

Let $C_1, \ldots, C_{\nu}$ be all such chains of indices.
Denote the length of some chain $C$ by $k(C)$, and the chain's $j$-th index by $i^C_j$.
Let $O$ be all the indices that are not included in any chain.
Then, for any $i\in O$, a similar equation to \cref{eq:GreedyChainStart} holds.
Overall:

\begin{align*}
\revenue(\alloc' | \adversary)
&
=
\sum_{i=0}^{\infty} \discount^i\fee_{\alloc'}(\adversary(i) \cup \pending_{\alloc'}(i))
\\&
=
\sum_{s=1}^{\nu} \sum_{j=0}^{k(C_s) - 1} \discount^{i^{C_s}_j}\fee(\alloc'(\adversary(i^{C_s}_j)\cup \pending_{\alloc'}(i^{C_s}_j)))
\\&
+
\sum_{s=1}^{\nu}  \discount^{i^{C_s}_{k(C_s)}}\fee(\alloc'(\adversary(i^{C_s}_{k(C_s)})\cup \pending_{\alloc'}(i^{C_s}_{k(C_s)})))
\\&
+
\sum_{i\in O} \discount^{i}\fee(\alloc'(\adversary(i)\cup \pending_{\alloc'}(i)))
\\&
\leq
\sum_{s=1}^{\nu} \discount \sum_{j=1}^{k(C_s)} \discount^{i^{C_s}_j}\fee(Greedy(\adversary(i^{C_s}_j)\cup \pending_{Greedy}(i^{C_s}_j)))
\\&
+
\sum_{s=1}^{\nu} \discount^{i^{C_s}_{k(C_s)}}\fee(Greedy(\adversary(i^{C_s}_{k(C_s)})\cup \pending_{Greedy}(i^{C_s}_{k(C_s)})))
\\&
+
\sum_{i\in O} \discount^{i}\fee(Greedy(\adversary(i)\cup \pending_{Greedy}(i)))
\\&
\leq
(1 + \discount) \sum_{s=1}^{\nu} \sum_{j=1}^{k(C_s)} \discount^{i^{C_s}_j}\fee(Greedy(\adversary(i^{C_s}_j)\cup \pending_{Greedy}(i^{C_s}_j)))
\\
& +
\sum_{i\in O} \discount^{i}\fee(Greedy(\adversary(i)\cup \pending_{Greedy}(i)))
\\&
\leq
(1 + \discount) \revenue(Greedy | \adversary)
\end{align*}
\end{proof}

\section{Proofs}
\label[appendix]{sec:Proofs}

\resDeterministicUpperBound*
\begin{proof}
We first give an overview of the proof. Let $a_1, \ldots, a_n$ be the series of fees chosen by the adversary $\adversary_n$ (like in our construction for the case $\discount < \frac{1}{\sqrt{2}}$ in the main text, we consider adversaries $\adversary_n$ that create $n$ transactions). The intuition from the  $\discount < \frac{1}{\sqrt{2}}$ case is that we should have $\lim_{n\rightarrow \infty} \frac{a_1}{1+\ell *a_1} \rightarrow  \frac{1}{a_1}$. 
However, when $\discount \imratio > 1$, the final terms are non-negligible, and in fact can be very large.
We thus need to adjust the series of fees $a_1, \ldots, a_n$ chosen by the adversary $\adversary_n$, so that each choice of the algorithm results in an equal competitive ratio.
Our main insight is that this is enough to prove $\frac{a_1}{1+\ell *a_1} \geq \frac{1}{a_1}$ for any $n$.
We do so by showing that the ratios $r_i = \frac{a_i}{a_{i-1}}$ are monotone increasing in $i$.

We next define the construction formally. 
Consider the set of adversaries $\Phi_n = \{\adversary_i\}_{i=1}^n \cup \{\adversary'_n\}$, which we define inductively:
\begin{align*}
    \adversary_1(0)
    &
    =
    \emptyset,
    \nonumber \qquad 
    \adversary_1(1)
    =
    \{(1,1),(2, a_1)\},
    \nonumber \qquad 
    \forall j>1, \adversary_1(j)
    =
    \emptyset
    \nonumber\\
    \forall n\geq i>1, \forall j < i, \adversary_i(j)
    &
    =
    \adversary_{i-1}(j),
    \nonumber \qquad 
    \adversary_i(i)
    =
    \{(1,a_{i-1}), (2,a_i),
    \nonumber \qquad
    \forall j > i, \adversary_i(j)
    =
    \emptyset
    \nonumber\\
    \forall j\leq n, \adversary'_n(j)
    &
    =
    \adversary_{n}(j),
    \nonumber \qquad
    \adversary'_n(n+1)
    =
    \{(1,a_n)\},
    \nonumber \qquad
    \forall j> n+1, \adversary'_n(j)
    =
    \emptyset
\end{align*}

Let $ALG$ be any deterministic allocation algorithm.
Let $k$ be the first index so that $ALG$ allocates a $TTL=2$ transaction when facing $\adversary'_n$, or undefined if there is no such index.
Then, if $k$ is defined, when facing $\adversary_k$, $ALG$ achieves competitive ratio of at most:
$$\frac{\sum_{i=1}^{k-1}\discount^i a_{i-1} + \discount^k a_k}{\sum_{i=1}^{k-1}\discount^i a_i + \discount^k a_{k-1} + \discount^{k+1}a_k}.$$
Otherwise, when facing $\adversary'_n$, $ALG$ achieves a competitive ratio of at most:
$$\frac{\sum_{i=1}^{n}\discount^i a_{i-1} + \discount^{n+1} a_n}{\sum_{i=1}^{n}\discount^i a_i + \discount^{n+1} a_{n}}.$$

We now find such $a_1, \ldots, a_n$ so that the above $n+1$ expressions are all equal.
Then, no matter the choice of $ALG$, the competitive ratio will be upper-bounded by this value $V$.
We show that when $n$ is large enough, this value tends to $\frac{1}{\frac{\discount}{2} + \frac{\sqrt{4+\discount^2}}{2}}$.

\subsubsection*{The solution always satisfies \texorpdfstring{$\frac{a_1}{1 + \discount a_1} \geq \frac{1}{a_1}$}{the inequality}}
There is always a unique solution $a_1 \geq 0, \ldots, a_n \geq 0$ to this system of equations.
Fix some $a_1$, then all other variables are determined by developing the equations and attaining the recursive relation:
\begin{equation}
\label{eq:RecursiveX}
     a_{i+1} = \frac{(1+V) (a_i - a_{i-1})}{\discount - V \discount^2}
\end{equation}
where $V = \frac{a_1}{1 + \discount a_1}, a_0 = 1$.
We have another equation, which we also attain by subtracting:
\begin{equation}
\label{eq:RecursiveX_n}
a_n = \frac{(1+V)a_{n-1}}{(1+V-\discount)}.
\end{equation}
The denominators of the above equations are always positive.
We prove that the numerator is always positive (and thus all $a_{i+1} \geq 0$) by inductively proving that $a_i \geq a_{i-1}$.
In fact, we prove a much stronger property: If we let $r_i = \frac{a_i}{a_{i-1}}$ (and let $r_1 = a_1, r_0 = 1$), we show that $r_i$ is monotonically increasing.
We rewrite \cref{eq:RecursiveX}:
\begin{equation}
r_{i+1} = \frac{(1+V) (1 - \frac{1}{r_i})}{\discount - V \discount^2}.
\end{equation}
If for some $i$, $r_i > r_{i-1}$, then $1 - \frac{1}{r_i} > 1 - \frac{1}{r_{i-1}}$, and so at the next step the numerator will be smaller and we will have $r_{i+1} > r_i$, and so on.
On the other hand, if for some $i$, $r_i \leq r_{i-1}$, then $r_i$ is weakly monotone decreasing.
For $r_n$, we recast \cref{eq:RecursiveX_n} as:
\begin{equation}
    r_n = \frac{1+V}{1+V-\discount}.
\end{equation}
Assume towards contradiction that $r_i$ is not monotonically increasing, then it must be that it is weakly monotonically decreasing and:
\begin{align*}
    \frac{1 + \discount a_1 + a_1}{1 + \discount a_1 - \discount - \discount^2 a_1}
    % &
    =
    \frac{1+V}{1+V-\discount}
    % \\&
    =
    r_n
    % \\&
    \leq
    r_1
    % \\&
    =
    a_1
    .
\end{align*}
This implies a lower bound of $a_1 \geq \frac{1}{1 - \discount} (1 + \frac{1}{\sqrt{\discount}})$.
We can also directly develop the formula for $a_2$ in terms of $a_1$ and $\discount$, which results in:
$a_2 = \frac{(1+\lambda)(a_1^2 - 1)}{\lambda} .$
Since it must be that $r_i$ is weakly monotonically decreasing, then $r_2 \leq r_1$, i.e., $a_2 \leq a_1^2$.
But, this is impossible given the formula for $a_2$ and the lower bound for $a_1$.
We conclude that $r_i$ is an increasing series, and in particular, $a_2 > a_1^2$.
This last inequality and the formula for $a_2$ are especially important as together they imply:
\begin{equation}
\label{eq:V_geq_x1}
\frac{a_1}{1 + \discount a_1} = V = \frac{1 + \discount a_2}{a_1 + \discount a_1 + \discount^2 a_2} \geq \frac{1}{a_1}.
\end{equation}
This holds if and only if $a_1 \geq \frac{\discount}{2} + \frac{\sqrt{4 + \discount^2}}{2}$.
Our argument also shows that $a_1 < \frac{1}{1 - \discount} (1 + \frac{1}{\sqrt{\discount}})$.

\subsubsection*{The solution satisfies \texorpdfstring{$\frac{a_1}{1 + \discount a_1} \rightarrow \frac{1}{a_1}$ as $n\rightarrow \infty$}{the inequality as n reaches infinity}}

It remains to show that for any $\epsilon > 0$, and any large enough $n$, it cannot hold that $a_1 \geq \frac{\discount}{2} + \frac{\sqrt{4 + \discount^2}}{2} + \epsilon$, and at the same time $V \geq \frac{1}{\imratio}$.
For this purpose, we represent the unique $a_1, \ldots, a_n$ solution as a sum of two exponentials.
By solving the quadratic equation:
$$\frac{\discount - V \discount^2}{1 + V} z^2 -z + 1 = 0,$$
with the roots $z_+ > z_-$, and choosing $A$ so that:
$a_1 = Az_+  + (1-A) z_-,$
we can ``roll'' the recursive relation of \cref{eq:RecursiveX} and write for any $i$,
\begin{equation}
a_i = Az_+^i + (1-A) z_-^i.
\end{equation}
We now note that a few facts that hold true for any $n$ when $a_1 \geq \imratio + \epsilon$:
\begin{enumerate}
    \item $z_+ \geq 1.01 \cdot z_-.$
    The two roots are positive whenever $V \geq \frac{1}{\imratio}$, and have:
$$
    \frac{z_+}{z_-}
    =
    \frac{\sqrt{4 \discount^2 V-4 \discount+V+1}+\sqrt{V+1}}{\sqrt{V+1}-\sqrt{4 \discount^2 V-4 \discount+V+1}}
    >
    1.01
    ,
$$
in the relevant regime.
Also,
$$
    z_+ \geq \frac{1+V}{2\discount}
    \geq
    \frac{1 + \imratio}{2\discount}
    \geq
    \frac{\imratio}{\discount}
    \geq
    1
.
$$

\item $A \geq \frac{1}{2}$.
We can derive a direct formula for $A$ in terms of $a_1, \discount$ when $a_1 \geq \imratio$:
$$\frac{1}{2} + \frac{1 + a_1 - l a_1}{2 \sqrt{\frac{1 + l (-4 + a_1) + a_1}{
 1 + l a_1}} (1 + l a_1) \sqrt{\frac{1 + a_1 + l a_1}{1 + l a_1}}},$$
which is composed of $\frac{1}{2}$ plus a positive term.

\item Finally, \cref{eq:RecursiveX_n} can be restated using the exponential representation as:
$$(1+V-\discount - (1 - V)z_+)A z_+^{n-1} + (1+V-\discount - (1 - V)z_-)(1-A)z_-^{n-1} = 0.$$

By our analysis in the first item, $z_+\geq 1$, so $(1+V-\discount - (1 - V)z_+) A \leq -\frac{\discount}{2}$, and:
$$(1+V-\discount - (1 - V)z_-)(1-A)z_-^{n-1} = -(1+V-\discount - (1 - V)z_+)A z_+^{n-1} \geq \frac{\discount}{2} z_+^{n-1}, $$
i.e.,
\[
\begin{split}
    \frac{1 + \frac{1}{1 - \discount} (1 + \frac{1}{\sqrt{\discount}})}{\discount}
    &
    >
    \frac{1 + a_1}{\discount}
    \nonumber\\&
    \geq
    \frac{1+V}{\discount}
    \nonumber\\&
    \geq
    \frac{1}{\discount} (1+V-\discount - (1 - V)z_-)
    \nonumber\\&
    \geq
    (1+V-\discount - (1 - V)z_-)(1-A) \frac{2}{\discount}
    \nonumber\\&
    \geq
    (\frac{z_+}{z_-})^{n-1}
    \nonumber\\&
    \geq
    1.01^{n-1}
    .
\end{split}
\]
This is a contradiction as the initial expression is a constant and the final expression increases exponentially with $n$.
Thus, for large enough $n$, either $V \leq \frac{1}{\imratio}$ (the upper bound we seek), or $a_1 < \imratio + \epsilon$, with arbitrarily low $\epsilon$, therefor $a_1 \rightarrow \imratio$, so $V \rightarrow \frac{1}{\imratio}$.
\end{enumerate}
\end{proof}

\resGoldenGreedy*
\Cref{res:GoldenGreedy} follows from the following claim. Setting $\imratio = \frac{1 + \discount \imratio}{\imratio}$ minimizes $\max \{\frac{1 + \discount \imratio}{\imratio}, \imratio\}$.
Solving the former, we get $\imratio = \frac{1}{2} (\discount + \sqrt{\discount^2 + 4})$, as required for the theorem:

\begin{restatable}[]{proposition}{resCubicBound}
\label{res:CubicBound}
For any $\imratio$, let $\Xi(\imratio) = \max \left \{\imratio, \frac{1 + \imratio\discount}{\imratio}, 1 + \discount^3\right \}$, then $R_{{\imratio}IB} \geq \frac{1}{\Xi(\imratio)}$.
\end{restatable}

\begin{proof}
Fix some arbitrary value of $\imratio$, and some adversary $\adversary$.
Consider an optimal allocation algorithm $\alloc'$ for this adversary as compared with $\alloc = {{\imratio}IB}$.
For a given index $i$, if both algorithms allocate a transaction with the same \gls{TTL} which arrives from $\adversary(i)$ (rather than from $\pending$), they must allocate the same one (the highest available for this \gls{TTL}).

If $\alloc'$ allocates at index $i$ a transaction that is available to ${{\imratio}IB}$ (i.e., it is in $\pending_{{{\imratio}IB}}(i)$), but ${{\imratio}IB}$ does not allocate it, then it must be that ${{\imratio}IB}$ allocates a transaction with a higher fee, or the best urgent transaction which pays a fee that is at least $\frac{1}{\imratio}$ as high as the fee paid by the one chosen by $\alloc'$.
Denote the set of indices where the latter happens by $I_0$.

If $\alloc'$ allocates a transaction at some index $i_0 \in I_0$ that is not available to ${{\imratio}IB}$ (i.e., it is not in $\pending_{{{\imratio}IB}}(i_0)$), then it must have been chosen by ${{\imratio}IB}$ at a previous index $i_1$.
If at index $i_1$, $\alloc'$ allocates a transaction that is not available to ${{\imratio}IB}$, then it must have been chosen by ${{\imratio}IB}$ at a previous index $i_2$, and so on.
Consider a series of indices $i_0, i_1, \dots$ that is constructed recursively in the same manner.
Since the index $i_0$ is a non-negative integer, and as the series is monotonically decreasing: $i_0 > i_1 > i_2 > \dots$, there must be some index $i_k$ where $\alloc'$ allocates a transaction that is available to ${{\imratio}IB}$.
Let $i_0 \define i$, and denote the recursively constructed ``chain'' of indices by $C = \{i_k, \ldots, i_0\}$.

First, notice that since for $0 \leq j \leq k - 1$ the allocation of ${{\imratio}IB}$ at index $i_{j+1}$ is the same as the allocation of $\alloc'$ at index $i_j$, we have, similarly to \cref{eq:GreedyChain}:
\begin{equation}
    \label{eq:BgreedyChain}
    \sum_{0 \leq j \leq k-1} \discount^{i_j}\fee(\alloc'(\adversary(i_j)\cup \pending_{\alloc'}(i_j)))
    % \\
    \leq \discount \sum_{1 \leq j \leq k} \discount^{i_j}\fee({{\imratio}IB}(\adversary(i_j)\cup \pending_{{{\imratio}IB}}(i_j))).
\end{equation}

We find a bound for $\fee(\alloc'(\adversary(i_{k})\cup \pending_{\alloc'}(i_{k})))$ by splitting into cases:
\begin{enumerate}
    \item If $\alloc'(\adversary(i_{k})\cup \pending_{\alloc'}(i_{k}))$ has $TTL = 1$, then since it is also available to ${{\imratio}IB}$, the best urgent transaction must pay a fee at least as high as paid by $\alloc'(\adversary(i_{k})\cup \pending_{\alloc'}(i_{k}))$.
    However, ${{\imratio}IB}$ allocates at step $i_k$ a higher $TTL$ transaction, as it is later chosen at $i_{k-1}$ by $\alloc'$.
    We conclude that ${{\imratio}IB}$ allocates the highest fee transaction at $i_k$, so it must satisfy:
    $$ \fee(\alloc'(\adversary(i_{k})\cup \pending_{\alloc'}(i_{k}))) \leq \frac{\fee({{\imratio}IB}(\adversary(i_k)\cup \pending_{{{\imratio}IB}}(i_k)))}{\imratio},$$
    and thus overall:
    \[
    \begin{split}
    \sum_{0 \leq j \leq k} \lambda^{i_j}\fee(\alloc'(\adversary(i_j)\cup \pending(i_j)))
    % \\
    & \leq
    \discount \sum_{1 \leq j \leq k} \discount^{i_j}\fee({{\imratio}IB}(\adversary(i_j)\cup \pending_{{{\imratio}IB}}(i_j)))
    \\&
    +
    \frac{1}{\imratio} \discount^{i_k}\fee({{\imratio}IB}(\adversary(i_j)\cup \pending_{{{\imratio}IB}}(i_k)))
    \\
    & \leq
    (\frac{1}{\imratio} + \discount) \sum_{1 \leq j \leq k} \discount^{i_j}\fee({{\imratio}IB}(\adversary(i_j)\cup \pending_{{{\imratio}IB}}(i_j)))
    \\
    & =
    \frac{1 + \imratio\discount}{\imratio} \sum_{1 \leq j \leq k} \discount^{i_j}\fee({{\imratio}IB}(\adversary(i_j)\cup \pending_{{{\imratio}IB}}(i_j))).
    \end{split}
    \]
    \item If $\alloc'(\adversary(i_{k})\cup \pending(i_{k}))$ has $TTL > 1$ and $i_{k-1} - i_k \geq 3$, then
    $\discount^{i_{k-1} - i_k} \leq \discount^3$, and so 
    \cref{eq:BgreedyChain} can be strengthened along the same lines as \cref{eq:GreedyChain}:
    \begin{equation}
    \label{eq:BgreedyChainStronger}
    \begin{split}
        & \sum_{0 \leq j \leq k-1} \discount^{i_j}\fee(\alloc'(\adversary(i_j)\cup \pending_{\alloc'}(i_j)))
        \leq
        \discount^3 \sum_{1 \leq j \leq k} \discount^{i_j}\fee({{\imratio}IB}(\adversary(i_j)\cup \pending_{{{\imratio}IB}}(i_j))) \\
    \end{split}
    \end{equation}
    This implies:
    \begin{align*}
    & \sum_{j=0}^{k - 1} \discount^{i_j}\fee(\alloc'(\adversary(i_j)\cup \pending_{\alloc'}(i_j)))
    % \\
    +
    \discount^{i_{k}}\fee(\alloc'(\adversary(i_{k})\cup \pending_{\alloc'}(i_{k})))
    \\
    & \leq
    (1 + \discount^3) \sum_{j=1}^{k} \discount^{i_j}\fee({{\imratio}IB}(\adversary(i_j)\cup \pending_{{{\imratio}IB}}(i_j)))
    \end{align*}
    \item If $\alloc'(\adversary(i_{k})\cup \pending(i_{k}))$ has $TTL > 1$ and $i_{k-1} = i_k + 1$ (the indices are consecutive), let:
    \begin{align*}
    \alpha
    &
    =
    \fee(\alloc'(\adversary(i_{k})\cup \pending(i_{k})))
    % ,
    \\
    \gamma
    &
    =
    \fee(\alloc'(\adversary(i_{k-1})\cup \pending(i_{k-1})))
    \end{align*}
    $\alloc'$ allocates $\alpha$ before $\gamma$ in the schedule, while it is possible for it to allocate $\gamma$ and then $\alpha$.
    We know that $\gamma \geq \alpha$ by our definition of the chain that ends when ${{\imratio}IB}$ allocates a higher-fee transaction then OPT (at index $i_k$).
    If $\gamma > \alpha$, then $\gamma + \discount \alpha > \alpha + \discount \gamma$, in contradiction to $\alloc'$ optimality against the adversary.
    If $\gamma = \alpha$, and since $\gamma$ has $TTL > 1$ it is available to ${{\imratio}IB}$ at step $i_{k-1}$.
    At each step of the chain, ${{\imratio}IB}$ allocates a non $TTL=1$ transaction, and so its choice at $i_{k-1}$ must be its highest-fee available transaction.
    But, then its choice is equal to that of $\alloc'$ at step $i_{k-1}$, and the chain should end there and not continue backward to $i_k$.

    \item If $\alloc'(\adversary(i_{k})\cup \pending(i_{k}))$ has $TTL > 1$ and $i_{k-1} = i_k + 2$ (i.e., there is exactly one transaction between the two indices), then similarly to before, its $TTL$ is at most $2$, or $\alloc'$ could better rearrange its choice.
    Since $1 < TTL \leq 2$, we conclude that $TTL = 2$.
    As in the previous case, denote the allocation of $\alloc'$ at $i_k$ as $\alpha$, the allocation of $\alloc'$ at $i_{k-1}$ as $\gamma$, and also denote its allocation at $i_k + 1$ (the in-between step) as $\beta$.
    We separate to several subcases.

    \begin{itemize}
    \item If $\alpha$ is not available to ${{\imratio}IB}$ at step $i_k + 1$, then it must have been chosen at the latest at some step $i \leq i_k - 1$.
    Assume that $\alpha$ is larger than OPT's allocated transaction at the step where ${{\imratio}IB}$ allocates it.
    Then it is not part of a chain (if it was part of a chain and larger than OPT's allocated transaction it would be the same chain as $i_k$, but that chain ends at $i_k$).
    We can thus ``attach'' it to the analysis of the chain without interfering in another chain's analysis.
    Denote OPT's allocated transaction at step $i$ as $s$.
    Since all the elements of $OPT$ in the chain besides $s, \alpha,$ and $\gamma$ were also chosen by ${{\imratio}IB}$ at an earlier step, they can only increase the competitive ratio, and so we can ignore them for the analysis.
    We are left with a ratio which is at most $\frac{\alpha + \discount^{i_k - i} \cdot \gamma}{\gamma + \discount^{i_k - i} \cdot \alpha + \discount^{2 + i_k - i} \cdot \gamma}$
    This is minimized with $i = i_k - 1, s = \gamma = \alpha$, yielding: $\frac{1 + \discount}{1 + \discount + \discount^3} \geq \frac{1}{1 + \discount^3}$.

    \item If $\alpha$ is not available to ${{\imratio}IB}$ at step $i_k + 1$, then it must have been chosen at the latest at some step $i \leq i_k - 1$.
    Assume that $\alpha$ is smaller than OPT's allocated transaction at the step where ${{\imratio}IB}$ allocates it, then the options are it is either a $TTL=1$ transaction as in case (1) (but then it cannot be later chosen by OPT, so this case is impossible), or it is at index $i_0$ of a chain.
    Notice that in our analysis of chains we always ignore ${{\imratio}IB}$'s chosen transaction at step $i_0$, since it only increases the competitive ratio and is not required for attaining the bounds.
    Thus, we can append it to the chain that we're analyzing without affecting the analysis of the chain it is a part of.
    We get a strict domination of ${{\imratio}IB}$ performance in our chain: Every transaction chosen by OPT is preceded by the same transaction chosen by ${{\imratio}IB}$, which yields a ratio of over $1$.
    \item If $\beta$ comes from the pending, i.e., is already available to $\alloc'$ at $i_k$, then $\beta \leq \alpha$ (by the same rearrangement argument we previously used).
    We know that $\alpha$ is available to ${{\imratio}IB}$ at step $i_k + 1$.
    If $\alpha$ or a higher $TTL = 1$ transaction is chosen by ${{\imratio}IB}$ at this step, then step $i_k + 1$ is not part of some chain (as ${{\imratio}IB}$ choices in a chain are always with $TTL > 1$), and so we can ``attach'' it to the analysis of the chain without interfering in another chain's analysis.
    Let:
    $$w = \frac{1}{\discount^3} \sum_{0 \leq j \leq k-2} \lambda^{i_j}\fee(\alloc'(\adversary(i_j)\cup \pending(i_j))).$$
    Then, over the path of the chain ${{\imratio}IB}$ gets at least:
    $$ \discount^{i_k} \cdot \gamma + \discount^{i_k + 1} \alpha + \discount^2 w,$$
    Moreover, $\alloc'$ gets at most:
    $$ \discount^{i_k} \cdot \alpha + \discount^{i_k + 1} \beta + \discount^{i_{k-1}}\cdot \gamma + \discount^3 w .$$
    With the constraint that $\gamma \geq \alpha \geq \beta, w \geq 0$, dividing the first expression by the latter yields a ratio of at least:
    $$\frac{\gamma + \discount \alpha + \discount^2 \cdot w}{\alpha + \discount \beta + \discount^2 \gamma + \discount^3 \cdot w}.$$
    This expression is minimized with $\gamma = \alpha = \beta$ and $w = 0$, where it becomes:
    $$
    \frac{1 + \discount}{1 + \discount + \discount^2} \geq  \min\{ \frac{1}{1 + \discount^3}, \frac{1}{\imratio}\}
    .
    $$

    \item If $\beta$ comes from the pending, i.e., is already available to $\alloc'$ at $i_k$, then $\beta \leq \alpha$ (by the same rearrangement argument we previously used).
    We know that $\alpha$ is available to ${{\imratio}IB}$ at step $i_k + 1$.
    If no $TTL = 1$ transaction is chosen by ${{\imratio}IB}$ at this step, then the highest-fee transaction $M$ with fee at least $\imratio \cdot \alpha$ is chosen by it, where recall that $\alpha \geq \beta$, the allocated transaction of $\alloc'$ at this step.
    Similarly to the previous step, this would guarantee a ratio better than $\frac{1}{1+\discount^3}$.
    However, in this case, it is possible that $i_k + 1$ is itself part of a chain, and we need to avoid double accounting of it.
    Notice however that in this case $i_k + 1$ is the earliest part of this chain, and the chain is of the type characterized in the first case (1), and since $i_k + 2 = i_{k-1}$ is part of the original chain, then the second-earliest link of the $i_k + 1$ chain is at least $i_k + 3$, which strengthens the guarantee of case (1) to $\frac{\imratio}{1 + \imratio \lambda^2}$.
    Joining the two chains together, and letting $w = \frac{1}{\discount^3} \sum_{0 \leq j \leq k-2} \lambda^{i_j}\fee(\alloc'(\adversary(i_j)\cup \pending(i_j)))$, we have that ${{\imratio}IB}$ gets at least $\discount^{i_k} \cdot \gamma + \discount^{i_k + 1}\cdot M  + \discount^2 w$, while $\alloc'$ gets at most $\discount^{i_k} \cdot x + \discount^{i_k + 1} \frac{1 + \discount^2 \imratio}{\imratio} \cdot M + \discount^{i_{k-1}}\cdot z + \discount^3 w $, where dividing the first expression by the latter yields a ratio of at least $$\frac{\gamma + \discount M + \discount^2 \cdot w}{\alpha + \discount \frac{1 + \discount^2 \imratio}{\imratio} \cdot M + \discount^2 \gamma + \discount^3 \cdot w},$$
    with the constraints $M \geq \imratio \cdot \alpha, \gamma \geq \alpha \geq \beta, w \geq 0$.
    The respective coefficients of $w, M, \gamma$ in the numerator and denominator yield a ratio that is at least as high as previous guarantees, and so the expression can only yield a worst guarantee if $w, M, \gamma$ are minimized, i.e., when $M = \imratio \alpha, w = 0, \gamma = \alpha$.
    The expression is then: $$\frac{1 + \discount \imratio}{1 + \discount (1 + \discount^2 \imratio) + \discount^2} = \frac{1 + \discount \imratio}{1 + \discount + \discount^2 +  \discount^3 \imratio} \geq \min\{ \frac{1}{1 + \discount^3}, \frac{1}{\imratio}\}.$$

    \item If $\beta$ does not come from $\pending$, then we cannot use a rearrangement argument to show that $\beta \leq \alpha$.
    Also, $\beta$ is available both to $\alloc'$ and ${{\imratio}IB}$ at step $i_k + 1$.
    If either $\beta \leq \alpha$, or that its $TTL$ at step $i_k + 1$ is $1$, the arguments of the previous subcases still hold, and so we consider only $\beta > \alpha$ with $TTL > 1$ (which also implies by a rearrangement argument that $\beta \geq \gamma$, where we recall that $\gamma$ is the allocated transaction of $\alloc'$ at $i_{k-1}$).
    If $\beta$ is not chosen at step $i_k + 1$, and since we may assume its TTL at this step is at least $2$, it will then be available to ${{\imratio}IB}$ at step $i_{k-1} = i_k + 2$, which contradicts that $i_{k-1}$ is part of the chain and ${{\imratio}IB}$ allocates its highest-fee available transaction there, as it allocates a transaction lower than $\gamma$ (and so lower than $\beta$).
    We conclude that ${{\imratio}IB}$ must allocate $\beta$ at step $i_k + 1$.

    If ${{\imratio}IB}$ allocates $\beta$, then we have that the ratio is at least $\frac{\gamma+\discount \beta}{\alpha + \discount \beta + \discount^2 \gamma}$ with $\beta \geq \gamma \geq \alpha$, and so this is at least as high as $\frac{1 + \discount}{1 + \discount + \discount^2} \geq \min\{ \frac{1}{1 + \discount^3}, \frac{1}{\imratio}\}$.
    \end{itemize}
\end{enumerate}

We consider chains $C$ where the opening pair falls into cases $1$, $2$ and $4$, grouped by the index sets $\mathcal{C}_1, \mathcal{C}_2$ and $\mathcal{C}_4$ respectively.
In all other cases, ${{\imratio}IB}$ allocates a transaction with a fee higher than $\alloc'$.
Let $k(C)$ be the length of some chain $C$, and $i^C_j$ the $j$-th index of a chain $C$.
Let $O$ denote the set of all such remaining indices.
We conclude that:

\begin{align*}
    \max_{\alloc'} \revenue(\alloc' | \adversary)
    & =
    \sum_{i\in I_0} \discount^i \alloc'(\adversary(i) \cup \pending_{\alloc'}(i))
    \nonumber \\
    & + \sum_{1\leq s \leq |\mathcal{C}_1|} \sum_{j=0}^{k(C_s)} \discount^{i^{C_s}_j} \alloc'(\adversary(i^{C_s}_j) \cup \pending_{\alloc'}(i^{C_s}_j)) \\
    & + \sum_{1 \leq s' \leq |\mathcal{C}_2|} \sum_{j=0}^{k(C_{s'})} \discount^{i^{C_{s'}}_j}  \alloc'(\adversary(i^{C_{s'}}_j) \cup \pending_{\alloc'}(i^{C_{s'}}_j)) \\
    & + \sum_{1 \leq s' \leq |\mathcal{C}_4|} \sum_{j=0}^{k(C_{s'})} \discount^{i^{C_{s'}}_j}  \alloc'(\adversary(i^{C_{s'}}_j) \cup \pending_{\alloc'}(i^{C_{s'}}_j)) \\
    & + \sum_{i\in O}  \discount^i \alloc'(\adversary(i) \cup \pending_{\alloc'}(i))
    \nonumber \\
    & \leq \imratio \sum_{i\in I_0} \discount^i {{\imratio}IB}(\adversary(i) \cup \pending_{{{\imratio}IB}}(i))
    \nonumber \\
    & + \frac{1 + \imratio\discount}{\imratio} \sum_{1\leq s \leq |\mathcal{C}_1|} \sum_{j=0}^{k(C_s)} \discount^{i^{C_s}_j} {{\imratio}IB}(\adversary(i^{C_s}_j) \cup \pending_{{{\imratio}IB}}(i^{C_s}_j)) \\
    & + (1 + \discount^3) \sum_{1\leq s' \leq |\mathcal{C}_2|} \sum_{j=0}^{k(C_{s'})} \discount^{i^{C_{s'}}_j} {{\imratio}IB}(\adversary(i^{C_{s'}}_j) \cup \pending_{{{\imratio}IB}}(i^{C_{s'}}_j)) \\
    & + \max\{1 + \discount^3, \imratio\} \sum_{1\leq s' \leq |\mathcal{C}_4|} \sum_{j=0}^{k(C_{s'})} \discount^{i^{C_{s'}}_j} {{\imratio}IB}(\adversary(i^{C_{s'}}_j) \cup \pending_{{{\imratio}IB}}(i^{C_{s'}}_j)) \\
    & + \sum_{i\in O} \discount^i {{\imratio}IB}(\adversary(i) \cup \pending_{{{\imratio}IB}}(i)) \\
    & \leq \max \{\imratio, \frac{1 + \discount \imratio}{\imratio}, 1+\discount^3\} \revenue({{\imratio}IB} | \adversary)
\end{align*}

\end{proof}

\resGoldenUpper*

\Cref{res:GoldenUpper} follows from the following technical result, when setting $\imratio = \frac{1}{2} (\discount + \sqrt{\discount^2 + 4})$:

\begin{restatable}[]{proposition}{resGoldenLower}
\label{res:GoldenLower}
For any $\imratio$, let $\Xi(\imratio) = \max \{\imratio, \frac{1 + \discount + \discount^2 \imratio + \discount^3}{1+\discount \imratio}, \frac{1 + \imratio\discount}{\imratio}, \max_{n\geq 1} \frac{\sum_{i=0}^{2n} \discount^i}{\sum_{i=0}^{n+1} \discount^i}\}$, then:
$$R_{{\imratio}IB} \leq \frac{1}{\Xi(\imratio)}.$$
\end{restatable}

\begin{proof}
Consider the adversary $\forall i\geq 0, \adversary(i) = \{(1,1), (2,\imratio-\epsilon)\}$, which is depicted in \cref{fig:GoldenUpperFirstAdversary}.
Given this adversary, ${{\imratio}IB}$ has a utility of $\revenue({{\imratio}IB} | \adversary) = 1 + \sum_{i=1}^{\infty} \discount^i$.
Furthermore, the allocation algorithm $\alloc'$ that allocates $(2,\imratio-\epsilon)$ at every step achieves a discounted utility equal to $\revenue(\alloc' | \adversary) = (\imratio-\epsilon) \cdot (1 + \sum_{i=1}^{\infty} \discount^i) = (\imratio-\epsilon) \revenue({{\imratio}IB} | \adversary)$.

\begin{figure}
    \centering
    \scalebox{1.0}{\begin{tikzpicture}[
  LeftArr/.style={shape=circle, minimum size=0.18cm, inner sep=0pt, draw=white, fill=white, label=center:$\leftarrow$},
  RightArr/.style={shape=circle, minimum size=0.18cm, inner sep=0pt, draw=white, fill=white, label=center:$\rightarrow$},
  Biased/.style={shape=rectangle, minimum size=0.2cm, inner sep=0pt, draw=mediumseagreen85168104, fill=mediumseagreen85168104},
  ALG/.style={shape=circle, minimum size=0.2cm, inner sep=0pt, draw=indianred1967882, fill=indianred1967882},
]
\tikzmath{
    \FirstStepNum   = 3;
    \LastStepNum    = 0;
    \StepNum        = \FirstStepNum + \LastStepNum + 1;
    \FirstTxNum     = 4;
    \LastTxNum      = 0;
    \TxNum          = \FirstTxNum + \LastTxNum;
    \MidTx          = -\TxNum/2-0.5;
}

% START Vertical lines
\node[] at (\StepNum/2, 0.5) {Step};
\foreach \x in {0, ..., \StepNum} {
    \draw[gray, dashed] (\x, -0.5) -- +(0, -\TxNum);
}
\foreach \x in {0, ..., {\FirstStepNum}} {
    \node[] at (\x+0.5, 0) {$\x$};
}
\node[] at (\FirstStepNum+1.5, 0) {...};
% END Vertical lines

% START TXs
\node[rotate=90] at (-3, \MidTx) {Transaction (\gls{TTL}, fee)};

\foreach \x in {0, 1} {
    \tikzmath{
        \y = -1-2*\x;
        \idxB = \x;
    };
    \draw[decorate, decoration={calligraphic brace, mirror}] (-1.5, {\y+0.25}) --  +(0, -1.5) node[left, midway]{$\adversary(\idxB)$};
    \node[] at (-0.75, \y) {$(1, 1)$};
    \draw[<->] (\x, \y) -- +(1, 0);
    \node[] at (-0.75, \y-1) {$(2, \imratio-\epsilon)$};
    \draw[<->] (\x, \y-1) -- +(2, 0);
}

\node[] at (-0.75, -4.4) {$\vdots$};
% END TXs

% START Allocations
% Step 0
\node[ALG] at (0.5, -1) {};
\node[Biased] at (0.5, -2) {};

% Step 1
\node[ALG] at (1.5, -3) {};
\node[Biased] at (1.5, -4) {};

% START Legend
\matrix[draw, right] at (\StepNum+1.5, \MidTx) {
    \node[LeftArr, label=right:Broadcast] {}; \\
    \node[RightArr, label=right:Expiration] {}; \\
    \node[ALG, label=right:$\alloc'$] {}; \\
    \node[Biased, label=right:${{\imratio}IB}$] {}; \\
};
% END Legend
\end{tikzpicture}}
    \caption{The first adversary used in the proof of \cref{res:GoldenLower}.}
    \label{fig:GoldenUpperFirstAdversary}
\end{figure}

Consider the adversary $\adversary(0) = \emptyset, \adversary(1) = \{(1,1),(2,\imratio + \epsilon)\}, \forall i\geq 2: \adversary(i) = \emptyset$.
Then the allocation algorithm $\alloc'$ that allocates $(1,1)$ at round $1$ and $(2,\imratio + \epsilon)$ at round $2$ obtains a utility of $\revenue(\alloc' | \adversary) = \cdot (\discount + (\imratio + \epsilon)\discount^2) = \frac{1 + (\imratio + \epsilon)\discount}{\imratio + \epsilon} \cdot \discount (\imratio + \epsilon) = \frac{1 + (\imratio + \epsilon)\discount}{\imratio + \epsilon}  \revenue({{\imratio}IB} | \adversary)$.

Notice that if we allocate $\imratio = \frac{1}{2} (\discount + \sqrt{\discount^2 + 4})$, the above two bounds also follow from the general deterministic upper bound of \cref{res:DeterministicUpperBound}.

Consider the adversary $\adversary_n$ that has $\adversary_n(0) = \emptyset, \forall 1 \leq i \leq n, \adversary_n(i) = \{(n+2, 1+\epsilon), (n + 2 -i, 1)\}, \forall i \geq n+1, \adversary_n(i) = \emptyset$.
Then the allocation algorithm $\alloc'$ that allocates the \emph{lowest} fee transaction at each round has $\revenue(\alloc' | \adversary_n) = \left(\discount \sum_{i=1}^n \discount^{i-1} + \sum_{i=n+1}^{2n} \discount^{i-1} (1 + \epsilon)\right)$, while $\revenue({{\imratio}IB} | \adversary_n) = \left(\discount \sum_{i=1}^n \discount^{i-1} (1 + \epsilon) + \discount^{n+1}\right)$.

Consider the adversary $\adversary(0) = \emptyset, \adversary(1) = \{(4,1), (1,\epsilon), (2, 1-\epsilon)\}, \adversary(2) = \{(2,\imratio+\epsilon), (1,1)\}$.
Then the algorithm $\alloc'$ that allocates $(2,1-\epsilon)$ at round $1$, $(1,1)$ at round $2$, $(2,\imratio+\epsilon)$ at round $3$ and $(4,1)$ at round $4$ has revenue arbitrarily close to:
$$\revenue(\alloc' | \adversary) = 1 + \discount + \discount^2 \imratio + \discount^3$$
On the other hand:
$$\revenue({{\imratio}IB} | \adversary) = 1 + \lambda \imratio .$$

\end{proof}

\resRandomizedUpperBound*

\begin{proof}
    Consider the set of fixed adversaries $\Phi_n = \{\adversary_i\}_{i=1}^n $. $\adversary_i$ are defined inductively: $\adversary_1$ with $\adversary_1(0) = \emptyset, \adversary_1(1) = \{(1,1),(2, \frac{2}{2-\discount})\}, \forall j>1, \adversary_1(j) = \emptyset$.
    Then, $\forall n\geq i>1, \forall j < i,\adversary_i(j) = \adversary_{i-1}(j), \adversary_i(i) = \{(1,(\frac{2}{2-\discount})^{i-1}), (2,(\frac{2}{2-\discount})^i), \forall j > i, \adversary_i(j) = \emptyset$.
    Due to the ``nesting'' structure of the construction, $ALG$'s performance against adversaries in $\Phi_n$ is defined by the series of probabilities $p_1, \ldots, p_n$, where $p_i$ is the probability that $ALG$ allocates a $TTL=1$ transaction at step $i$.
    
    We now use $\Phi_n$ to define our adaptive adversary $\adversary^n_{adapt}$, and compare the performance of an algorithm we call ADV to this adversary's response.
    At any step $i$, if $p_i > \frac{1}{2}$, ADV allocates the $TTL=2$ transaction, and otherwise if $p_i \leq \frac{1}{2}$, allocates the $TTL=1$ one.
    In the latter case, if $ALG$ allocates the $TTL=2$ transaction (which happens w.p. $1 - p_i$), the adversary does not add transactions to the pending set, i.e., the adversary is finalized to be $\adversary_i$, and at step $i+1$ ADV allocates the remaining $TTL=1$ transaction.
    If no such step $i$ happens before $i=n$, then the adversary terminates with $\adversary_n$.
    If $p_n > \frac{1}{2}$ and ADV allocates the $TTL=2$ transaction at step $n$ while $ALG$ allocates the $TTL=1$ transaction, the adversary adds a transaction $(1, (\frac{2}{2-\discount})^i)$ at step $n+1$, and ADV allocates it.
    
    We now compare $ALG$ and $ADV$'s performance in different steps along the adversarial arrival, separating the steps before $n$ and the last two steps.

    \subsubsection*{Step \texorpdfstring{$i<n$}{i < n}}
    Examine each one separately:
    \begin{itemize}
        \item The expected performance of $ALG$ is:
        $(\frac{2}{2-\discount})^{i-1}\left(p_i + (1-p_i) \frac{2}{2-\discount}\right).$
    
        \item The expected performance of $ADV$ at steps $i$, and possibly $i+1$ if step $i$ is a terminal step, when $i < n$ and $p_i \leq \frac{1}{2}$ is:
        $(\frac{2}{2-\discount})^{i-1} \left(1 + (1-p_i)\discount \frac{2}{2-\discount} \right)$ (we amortize the added utility of step $i+1$ into the utility of step $i$).
        Notice that this amortization of considering the $i+1$ is only relevant for $ADV$, as $ALG$ in such case necessarily has no transactions remaining to allocate from at step $i+1$.

        \item The expected performance of $ADV$ at step $i < n$ if $p_i > \frac{1}{2}$ is:
        $(\frac{2}{2-\discount})^i.$
    \end{itemize}

    For $i < n$, we first consider $p_i > \frac{1}{2}$:
    \[
    \begin{split}
        & E\left[\fee_{ADV}(\adversary(i) \cup \pending_{ADV}(i))\right]
 %       \\&
        =
        (\frac{2}{2-\discount})^i
 %       \\&
        =
        (\frac{2}{2-\discount})^{i-1} \cdot \frac{2}{2 - \discount}
%        \\&
        =
        \frac{2}{2-\discount}^{i-1} (\frac{1}{2} + \frac{1}{2-\discount} ) \frac{\frac{2}{2 - \discount}}{\frac{1}{2} + \frac{1}{2-\discount}}
        \\&
        =
        \frac{2}{2-\discount}^{i-1} \left((p_i + (1-p_i)\frac{2}{2-\discount} ) + (\frac{1}{2} - p_i + (\frac{1}{2} - (1- p_i)) \frac{2}{2-\discount})\right) \frac{\frac{2}{2 - \discount}}{\frac{1}{2} + \frac{1}{2-\discount}}
        \\&
        =
        \frac{4}{4- \discount} \cdot E\left[\fee_{ALG}(\adversary(i) \cup \pending_{ALG}(i))\right] + (\frac{2}{2-\discount})^{i-1} (p_i - \frac{1}{2}) \frac{\discount}{2-\discount} \frac{4}{4- \discount}
        .
    \end{split}
    \]
    For $p_i \leq \frac{1}{2}$, we have:
    \[
    \begin{split}
        & E\left[\fee_{ADV}(\adversary(i) \cup \pending_{ADV}(i))\right] + (1-p_i)E\left[\fee_{ADV}(\adversary(i+1) \cup \pending_{ADV}(i+1))\right]
        \\&
        =
        (\frac{2}{2-\discount})^{i-1} \left(1 + (1-p_i)\discount \frac{2}{2-\discount} \right)
    \end{split}
    \]
    And then, one can calculate:
    \[
    \begin{split}
        (\frac{2}{2-\discount})^{i-1} \left(1 + (1-p_i)\discount \frac{2}{2-\discount} \right)
        &
        =
        (\frac{2}{2-\discount})^{i-1}\left(p_i + (1-p_i) \frac{2}{2-\discount}\right) \frac{1 + (1-p_i)\discount \frac{2}{2-\discount}}{p_i + (1-p_i) \frac{2}{2-\discount}}
        \\&
        =
        E\left[\fee_{ALG}(\adversary(i) \cup \pending_{ALG}(i))\right]\cdot \frac{1 + (1-p_i)\discount \frac{2}{2-\discount}}{p_i + (1-p_i) \frac{2}{2-\discount}}
        \\&
        \geq
        \frac{4}{4- \discount} \cdot E\left[\fee_{ALG}(\adversary(i) \cup \pending_{ALG}(i))\right]
        ,
    \end{split}
    \]
    where the last transition is since for any $0 \leq \discount \leq 1$, the expression
    $\frac{1 + (1-p_i)\discount \frac{2}{2-\discount}}{p_i + (1-p_i) \frac{2}{2-\discount}}$
    under the constraint $0 \leq p_i \leq \frac{1}{2}$ is minimized with $p_i = \frac{1}{2}$, where it takes the form $\frac{4}{4- \discount}$.
    
    We conclude from the two cases that:
    \[
    \begin{split}
        E^{Amortized}\left[\fee_{ADV}(\adversary(i) \cup \pending_{ADV}(i))\right]
        &
        \geq
        \frac{4}{4- \discount} \cdot E\left[\fee_{ALG}(\adversary(i) \cup \pending_{ALG}(i))\right]
        \\&
        +
        \max \{0, (\frac{2}{2-\discount})^{i-1} (p_i - \frac{1}{2}) \cdot \frac{\discount}{2-\discount} \cdot \frac{4}{4- \discount}\}
    \end{split}
    \]

    We now analyze steps $n, n+1$, starting with the expected performance at these steps.
    \begin{itemize}
        \item The expected performance of $ALG$ at these steps is:
        $$(\frac{2}{2-\discount})^{n-1}\left(p_n(1 + \discount \cdot \frac{2}{2-\discount}) + (1-p_n) \frac{2}{2-\discount}\right)$$

        \item If $p_n \leq \frac{1}{2}$, the expected performance of $ADV$ at these steps is:
        $$(\frac{2}{2-\discount})^{n-1} \left( 1 + \discount \cdot \frac{2}{2-\discount} \right)$$

        \item If $p_n > \frac{1}{2}$, the expected performance of $ADV$ at these steps is:
        $$(\frac{2}{2-\discount})^{n-1} \left( \frac{2}{2-\discount} + \discount \cdot \frac{2}{2-\discount} \right)$$
    \end{itemize}

    Let $f(p_n) = \begin{cases} (2-p_n) & p_n > \frac{1}{2} \\
    (1 - p_n) & p_n \leq \frac{1}{2}\end{cases}$.
    One can show that:
    \begin{equation}
    \label{eq:ADV_ALG_n}
    \begin{split}
    & E\left[\fee_{ADV}(\adversary(n) \cup \pending_{ADV}(n))\right] + \discount E\left[\fee_{ADV}(\adversary(n+1) \cup \pending_{ADV}(n+1))\right] \geq \\
    & E\left[\fee_{ALG}(\adversary(n) \cup \pending_{ALG}(n))\right] + \discount E\left[\fee_{ALG}(\adversary(n+1) \cup \pending_{ALG}(n+1))\right] + \discount^n (\frac{2}{2-\discount})^{n-1} \frac{\discount}{2-\discount} \cdot f(p_n)
    .
    \end{split}
    \end{equation}

    It also holds that \begin{equation}
    \label{eq:fpn_pn}
    f(p_n) - p_n = \begin{cases} (2-2p_n) & p_n > \frac{1}{2} \\
    (1 - 2p_n) & p_n \leq \frac{1}{2}\end{cases} \geq 0.\end{equation}
    
    We next prove by backward induction that:
    $$u(ADV | \adversary^n_{adapt})
    \geq
    \frac{4}{4-\discount} u(ALG | \adversary^n_{adapt}) - \frac{2\discount^2}{(4 - \discount)(2-\discount)}
    .
    $$

    We prove so by the inductive claim:
    \[
    \begin{split}
        & E^{Amortized}\left[\sum_{i=k}^{n+1} \discount^i \fee_{ADV}(\adversary(i) + \pending(i))\right]
        \\
        & \geq
        \frac{4}{4-\discount} E\left[\sum_{i=k}^{n+1} \discount^i \fee_{ALG}(\adversary(i) + \pending(i))\right] - \frac{\discount}{4 - \discount} (\frac{2\discount}{2-\discount})^k
    \end{split}
    \]

    As the base case, consider $k=n$.
    Then,
    \[
    \begin{split}
        & E\left[\sum_{i=n}^{n+1} \discount^i \fee_{ADV}(\adversary(i) + \pending(i))\right]
        \\&
        \stackrel{\text{\cref{eq:ADV_ALG_n}}}{\geq}
         E\left[\sum_{i=n}^{n+1} \discount^i \fee_{ALG}(\adversary(i) + \pending(i))\right]
        +
        \discount^n (\frac{2}{2-\discount})^{n-1} \frac{\discount}{2-\discount} f(p_n)
        \\&
        =
        \frac{4}{4-\discount} E\left[\sum_{i=n}^{n+1} \discount^i \fee_{ALG}(\adversary(i) + \pending(i))\right]
        +
        \discount^n (\frac{2}{2-\discount})^{n-1} \frac{\discount}{2-\discount} f(p_n)
        \\&
        - \frac{\discount}{4-\discount} \discount^n \left( E\left[\fee_{ALG}(\adversary(n) + \pending(n))\right]
        +
        \discount E\left[\fee_{ALG}(\adversary(n+1) + \pending(n+1 ))\right]\right)
        \\&
        \geq
        \frac{4}{4-\discount}  E\left[\sum_{i=n}^{n+1} \discount^i \fee_{ALG}(\adversary(i) + \pending(i))\right]
        +
        \discount^n (\frac{2}{2-\discount})^{n-1} \frac{\discount}{2-\discount} f(p_n)
        \\&
        - \frac{\discount}{4 - \discount} \discount^n (\frac{2}{2-\discount})^{n-1} (\frac{2}{2-\discount} + p_n \discount \frac{2}{2-\discount})
        \\&
        \geq
        \frac{4}{4-\discount} E\left[\sum_{i=n}^{n+1} \discount^i \fee_{ALG}(\adversary(i) + \pending(i))\right]
        +
        \discount^n (\frac{2}{2-\discount})^{n-1} \frac{\discount}{2-\discount} (f(p_n)-p_n)
        \\&
        -
        \frac{\discount}{4 - \discount} \discount^n (\frac{2}{2-\discount})^{n-1} \frac{2}{2-\discount}
        \\&
        \stackrel{\text{\cref{eq:fpn_pn}}}{\geq}
        \frac{4}{4-\discount}  E\left[\sum_{i=n}^{n+1} \discount^i \fee_{ALG}(\adversary(i) + \pending(i))\right]
        -
        \frac{\discount}{4 - \discount}  (\frac{2\discount}{2-\discount})^n
        .
    \end{split}
    \]
    For the inductive step:
    \[
    \begin{split}
        E^{Amortized}\left[\sum_{i=k}^{n+1} \discount^i \fee_{ADV}(\adversary(i) + \pending(i))\right]
        &
        =
        E^{Amortized}\left[\discount^k \fee_{ADV}(\adversary(k) + \pending(k))\right]
        \\&
        +
        p_k \cdot E^{Amortized}\left[\sum_{i=k+1}^{n+1} \discount^i \fee_{ADV}(\adversary(i) + \pending(i))\right]
    \end{split}
    \]
    We can then bound this from below:
    \[
    \begin{split}
        & E^{Amortized}\left[\sum_{i=k}^{n+1} \discount^i \fee_{ADV}(\adversary(i) + \pending(i))\right] \\
        &
        \geq
        \frac{4}{4- \discount} \cdot E\left[\fee_{ALG}(\adversary(k) \cup \pending_{ALG}(k))\right]
        +
        \max \{0, (\frac{2}{2-\discount})^{k-1} (p_k - \frac{1}{2}) \frac{\discount}{2-\discount} \frac{4}{4- \discount}\}
        \\&
        +
        p_k \cdot \frac{4}{4- \discount} E\left[\sum_{i=k+1}^{n+1} \discount^i \fee_{ALG}(\adversary(i) + \pending(i))\right]
        -
        p_k \frac{\discount}{4 - \discount}  (\frac{2\discount}{2-\discount})^{k+1}
        \\&
        =
        \frac{4}{4- \discount} E\left[\sum_{i=k}^{n+1} \discount^i \fee_{ALG}(\adversary(i) + \pending(i))\right]
        +
        \max \{0, \frac{2}{2-\discount}^{k-1} (p_k - \frac{1}{2}) \frac{\discount}{2-\discount} \frac{4}{4- \discount}\}
        -
        p_k \frac{\discount}{4 - \discount}  (\frac{2\discount}{2-\discount})^{k+1}
    \end{split}
    \]
    We thus need to show that:
    $$\max \{0, (\frac{2}{2-\discount})^{k-1} (p_k - \frac{1}{2})\frac{\discount}{2-\discount}\frac{4}{4- \discount}\} - p_k \frac{\discount}{4 - \discount}  (\frac{2\discount}{2-\discount})^{k+1} \geq -\frac{\discount}{4 - \discount}  (\frac{2\discount}{2-\discount})^k. $$
    First, consider if $p_k \leq \frac{1}{2}$.
    Then,
    \[
    \begin{split}
        & \max \{0, (\frac{2}{2-\discount})^{k-1} (p_k - \frac{1}{2}) \frac{\discount}{2-\discount} \frac{4}{4- \discount}\} - p_k \frac{\discount}{4 - \discount}  (\frac{2\discount}{2-\discount})^{k+1} \\
        & 
        =
        - p_k \frac{\discount}{4 - \discount}  (\frac{2\discount}{2-\discount})^{k+1}
        \geq
        - \frac{1}{2}\frac{\discount}{4 - \discount} (\frac{2\discount}{2-\discount})^{k+1}
    \end{split}
    \]
    Thus, we need to show that $-\frac{1}{2} \frac{2\discount}{2-\discount} \geq -1$, i.e., $-\discount \geq \discount - 2$.
    This indeed holds, since $2 \geq 2\discount$.
    Otherwise, consider if $p_k > \frac{1}{2}$, then,
    \[
    \begin{split}
        &
        \max \{0, (\frac{2\discount}{2-\discount})^{k-1} (p_k - \frac{1}{2}) \frac{\discount}{2-\discount} \frac{4}{4- \discount}\} - p_k \frac{\discount}{4 - \discount} (\frac{2\discount}{2-\discount})^{k+1}
        \\&
        =
        (\frac{2\discount}{2-\discount})^k (p_k - \frac{1}{2})\frac{2\discount}{4- \discount}
        -
        p_k \frac{\discount}{4 - \discount}  (\frac{2\discount}{2-\discount})^{k+1}
        \\&
        \geq
        \frac{\discount}{4 - \discount}  (\frac{2\discount}{2-\discount})^k \left(2p_k - 1 -p_k \frac{2\discount}{2-\discount} \right)
    \end{split}
    \]
    It thus suffices to show $2p_k - 1 -p_k \frac{2\discount}{2-\discount} \geq -1$, which holds if and only if:
    $2 \geq \frac{2\discount}{2-\discount},$
    which holds since $4 \geq 4 \discount$.
    Consider $\lim_{n\rightarrow \infty} u(ALG | \adversary^n_{adapt})$.
    
    If $\lim_{n\rightarrow \infty} u(ALG | \adversary^n_{adapt}) = \infty$, then:
    \[
    \begin{split}
    \lim_{n\rightarrow \infty} \frac{u(ALG | \adversary^n_{adapt})}{u( ADV | \adversary^n_{adapt})}
    &
    \leq
    \lim_{n\rightarrow \infty} \frac{u(ALG | \adversary^n_{adapt})}{\frac{4}{4-\discount}u(ALG | \adversary^n_{adapt}) -  \frac{2\discount^2}{(4 - \discount)(2-\discount) }}
    =
    \frac{4-\discount}{4}.\\
    \end{split}
    \]
    Otherwise, if $\lim_{n\rightarrow \infty} u(ALG | \adversary^n_{adapt})$ is finite, the contribution of steps $n, n+1$ to the sum tends to zero.
    Thus, a similar argument for the negligence of the additive term follows.
\end{proof}
\end{document}